\documentclass[fleqn,usenatbib]{mnras}

\usepackage[T1]{fontenc}

\DeclareRobustCommand{\VAN}[3]{#2}
\let\VANthebibliography\thebibliography
\def\thebibliography{\DeclareRobustCommand{\VAN}[3]{##3}\VANthebibliography}

\usepackage{graphicx}	
\usepackage{amsmath}	
\usepackage{amssymb}	

\usepackage{soul}

\usepackage{newtxtext,newtxmath}

\newcommand{\cms}{\,$\text{cm}\,\text{s}^{-1}\,$}	
\newcommand{\ms}{\,$\text{m}\,\text{s}^{-1}\,$}	
\newcommand{\msnw}{\,$\text{m}\,\text{s}^{-1}$}	

\newcommand{\kms}{\,$\text{km}\,\text{s}^{-1}$}	

\title[MM-LSD]{Multi-Mask Least-Squares Deconvolution: Extracting RVs using tailored masks}

\author[Lienhard et al.]{
F. Lienhard,$^{1}$\thanks{E-mail: fl386@cam.ac.uk}
A. Mortier,$^{1,2}$
L. Buchhave,$^{3}$
A. Collier Cameron,$^{4}$
M. L\'opez-Morales,$^{5}$
A. Sozzetti,$^{6}$\newauthor
C. A. Watson,$^{7}$
R. Cosentino$^{8}$
\\
$^{1}$Astrophysics Group, Cavendish Laboratory, University of Cambridge, J.J. Thomson Avenue, Cambridge CB3 0HE, UK\\
$^{2}$Kavli Institute for Cosmology, University of Cambridge, Madingley Road, Cambridge CB3 0HA, UK\\
$^{3}$DTU Space, National Space Institute, Technical University of Denmark, Elektrovej 328, DK-2800 Kgs. Lyngby, Denmark\\
$^{4}$SUPA School of Physics and Astronomy, University of St Andrews, North Haugh, St Andrews KY16 9SS, UK\\
$^{5}$Center for Astrophysics | Harvard \& Smithsonian, 60 Garden Street, Cambridge, MA 02138, USA\\
$^{6}$INAF -- Osservatorio Astrofisico di Torino, via Osservatorio 20, I-10025 Pino Torinese, Italy\\
$^{7}$Astrophysics Research Centre, School of Mathematics and Physics, Queen's University Belfast, BT7 1NN, Belfast, UK\\
$^{8}$Fundaci\'on Galileo Galilei-INAF, Rambla Jos\'e Ana Fernandez P\'erez 7, E-38712 Bre\~na Baja, TF, Spain\\
}

\date{Accepted 2022 April 09. Received 2022 April 08; in original form 2021 October 25}

\pubyear{2015}

\begin{document}
\label{firstpage}
\pagerange{\pageref{firstpage}--\pageref{lastpage}}
\maketitle

\begin{abstract}
To push the radial velocity (RV) exoplanet detection threshold, it is crucial to find more reliable radial velocity extraction methods. The Least-Squares Deconvolution (LSD) technique has been used to infer the stellar magnetic flux from spectropolarimetric data for the past two decades. It relies on the
assumption that stellar absorption lines are similar in shape. Although this assumption is simplistic,
LSD provides a good model for intensity spectra and likewise an estimate for
their Doppler shift. We present the Multi-Mask Least-Squares Deconvolution (MM-LSD) RV extraction pipeline which extracts the radial velocity from two-dimensional echelle-order spectra using LSD with multiple tailored masks after continuum normalisation and telluric absorption line correction. The flexibility of LSD allows to exclude spectral lines or pixels at will, providing a means to exclude variable lines or pixels affected by instrumental problems. The MM-LSD pipeline was tested on HARPS-N data for the Sun and selected well-observed stars with 5.7 $<$ Vmag $<$ 12.6. For FGK-type stars with median signal-to-noise above 100, the pipeline delivered RV time series with on average 12 per cent lower scatter as compared to the HARPS-N RV extraction pipeline based on the Cross-Correlation Function technique. The MM-LSD pipeline may be used as a standalone RV code, or modified and extended to extract a proxy for the magnetic field strength.
\end{abstract}




\section{Introduction}

Since the discovery of the first planet orbiting a solar-type star, 51 Peg b \citep{peg51b1995}, the radial velocity (RV) community has made significant progress in instrumentation and data processing. In the past years, it became apparent that the data processing and modelling techniques are lagging behind the improvements in instrument design. At the current state, the instrumental precision and stability of instruments such as ESPRESSO \citep{Pepe_2021}, HARPS-N \citep{Cosentino_2012}, or EXPRES \citep{Jurgenson_2016,Blackman_2020} are about or below 1\ms. The detection of RV signals of this magnitude remains very challenging, to the extent that the detection or characterisation of an Earth twin (10\cms radial velocity semi-amplitude) is out of reach. New radial velocity extraction or postprocessing techniques are therefore key to the detection of Earth-like exoplanets and more accurate mass determinations of known planets.

The main obstacle to detecting RV signals of smaller planets is stellar activity. The latter encompasses a multitude of phenomena on stellar surfaces, such as starspots, faculae, plages, or granules. One effect in this context is the suppression of convective blueshift \citep[][]{Meunier_2010a} which significantly impacts the RV measurements for solar-like stars, as described in the following. The stellar photosphere is covered with granules, where hot gas is moving upward, and intergranular lanes, where the cooler gas is sinking back downward \citep[for a review see][]{Stein_2012}. Due to the temperature difference and the granules occupying more space, the upward moving gas dominates the emission causing a net blueshift for the observer measuring the integrated spectrum \citep[e.g.][]{Dravins_1981,Dravins_1982}. Since the magnetic field inhibits convection \citep[e.g.][]{Hanslmeier_1991} and evolves in time, the time-dependent and locally inhomogeneous suppression of convective motion induces RV signals.
Strong local magnetic fields can similarly produce starspots. 

There are two factors causing the light emanating from a starspot to differ from the light emitted from quiet surface regions. Firstly, the Doppler shift of the light emitted from a starspot differs from the light emitted from quiet surface regions due to the reduced convective flow. Secondly, the starspot's temperature is lower, making it appear dimmer. 
These two factors significantly contribute to the spectrums variability since a star is rotating and typically not seen perfectly pole-on and thus the brightness inhomogeneity travels from the blue-shifted to the red-shifted side of the star. In fact, the Doppler imaging technique takes advantage of this fact to map the brightness distribution on the stellar surface \citep[e.g.][]{Vogt_1983,Vogt_1987,Cameron_1994,Donati_1997_doradus,Crossfield_2014}. In the RV context, the variation of the rotationally broadened absorption lines can induce spurious RV signals \citep[e.g.][]{Queloz_2001}. The impact of starspots on RV measurements was studied in \citet[][]{Saar_1997,Desort_2007,Lagrange_2010}.

There are a number of other sources of astrophysical noise, including granulation \citep{Dravins_1982,Dumusque_2011,Meunier_2015,Cegla_2018}, super-granulation \citep{Rieutord_2010,Rincon_2018,Meunier_2019}, p-mode oscillations \citep{Mayor_2003,Chaplin_2019}, and surface magnetic activity features such as faculae and plages \citep{Meunier_2010a,Meunier_2010b} that all operate on different timescales and magnitudes.

A promising way forward to disentangle stellar and planetary signals in RV data is the unsigned hemispherically-averaged stellar magnetic flux which is known to correlate with solar RV variations \citep{Haywood_2016,Haywood_2020}. This flux can be measured by accounting for Zeeman splitting of spectral lines. 

Multiple authors have developed new RV extraction techniques \citep[e.g.][]{Anglada_Escude_2012,Dumusque_2018,Zechmeister_2018,Bedell_2019,Holzer_2020,Rajpaul_2020} with varying advantages and disadvantages. Nevertheless, the Cross-Correlation Function (CCF) technique \citep{Baranne_1996,Pepe_2002} is still the standard method employed to deduce a star’s radial velocity from its spectrum. The HARPS-N spectrograph, for instance, routinely applies the CCF technique on the extracted two-dimensional echelle order spectra (S2D). In this work, we present the Multi-Mask Least-Squares Deconvolution (MM-LSD) RV extraction pipeline\footnote{https://github.com/florian-lienhard/MM-LSD (available upon publication)} based on Least-Squares Deconvolution (LSD) providing a precise estimate of the stellar RV, average profiles of the absorption lines within each echelle order, and a simple model for the spectrum. We show that the RV scatter is generally lower for MM-LSD compared to the CCF technique, while the planetary signal is more prominent. The MM-LSD technique is flexible with regards to the used line mask and spectral model. In this paper, we describe the basic usage of our for extracting radial velocities. We will extend the pipeline to account for Zeeman splitting in a future iteration of this work. 

In Section \ref{s:Data}, we present the used data, such as the spectra and the input used for the mask generation. The theoretical framework employed to extract the radial velocity from the spectra is described in Section \ref{s:LSD_Background}.
The estimation and correction of the stellar continuum for the echelle order spectra is outlined in Section \ref{s:Continuum_correction}. The detailed workings of the RV extraction is presented in Section \ref{s:LSD_algo_description}. In Section \ref{ss:findparameters}, we describe how the hyperparameters for the LSD procedure are set. Next, we show and analyse the performance of the LSD RV extraction pipeline in Section \ref{s:results}. We conclude in Section \ref{s:conclusion_outlook} and discuss future work as well as potential applications of the code.

\section{Data} \label{s:Data}

In this work, we use HARPS-N data, but the described code is applicable to all high resolution spectra. Additionally to the HARPS-N data, we need some information about the stellar and telluric absorption lines' position and depth. These three sources of information are described in the following subsections.

\subsection{HARPS-N spectroscopy} \label{ss:hn_spectroscopy}

HARPS-N \citep{Cosentino_2012} is a pressure and temperature stabilised cross-dispersed echelle spectrograph installed at the Telescopio Nazionale Galileo in the Canary Islands. The instrument has been operational since 2012 producing spectra with a resolving power of R = 115,000 in the visible range from 383 to 690 nm over 69 spectral orders.
Since 2015, the HARPS-N solar telescope has additionally been observing the Sun for several hours on most days, producing disk-integrated spectra at a 5 minute cadence \citep{Cosentino_2014,Dumusque_2015,Phillips_2016,Collier_Cameron_2019,Dumusque_2021}. Here, we use the ESPRESSO pipeline products applied on HARPS-N spectra as described in \citet{Dumusque_2021}. More specifically, we make use of the two-dimensional echelle order spectra (S2D) and the one-dimensional spectra merged on to grid of wavelengths with uniform logarithmic spacing (S1D) produced by the DRS pipeline version 2.3.1 for the stellar spectra and version 2.2.2 for the solar spectra (public data). Note that both data products have been corrected for the blaze function and their wavelength solutions have been corrected for instrumental drift. The tested stars are listed in Table \ref{tab:starproperties}. These stars were selected for their diverse observational and intrinsic properties, such as the brightness, airmass, and the number of planets orbiting the respective star.

\subsection{VALD3} \label{ss:vald}

The mask used in the LSD procedure is based on \textit{a priori} absorption line information. For this work, we only require the absorption line wavelength and central depth. These can be, like is the case in the DRS pipeline, extracted from high SNR spectra of well-known stars similar to the target at hand. Since we want to retain the physics of the absorption lines without depending on the quality of a specific mask, we opt for retrieving the absorption line wavelength and depth from the Vienna Atomic Line Database (VALD3; \citealt{Ryabchikova_2015}).

{\sc MM-LSD} reads the standard VALD3 output from ``Extract Stellar'' in the ``Long format''. VALD3 needs the minimum depth of the absorption lines that we want to retrieve (1 per cent), the wavelength range of the instrument, as well as the stellar microturbulence, effective temperature ($T_{\text{eff}}$), and surface gravity ($\log g$) as input parameters. The chemical composition was set to solar values for all stars in this study, while the other parameters were obtained as outlined in Section \ref{s:results} and are listed in Table \ref{tab:starproperties}. The depth threshold of 1 per cent was chosen for computational efficiency and as a consequence of the marginal impact of shallower lines on the LSD model and the resulting RV.

\subsection{Stellar parameters} \label{ss:stellar_parameters}
In Table \ref{tab:starproperties}, we provide information about the test stars for this study, the included HARPS-N spectra, and the properties used to get the VALD3 line list: effective temperature, surface gravity, and microturbulence. These properties were derived using a co-added HARPS-N spectrum for each star. We employed a curve-of-growth method using a set of neutral and ionised iron lines. This method, ARES+MOOG\footnote{ARESv2: \url{http://www.astro.up.pt/~sousasag/ares/}; MOOG 2017: \url{http://www.as.utexas.edu/~chris/moog.html}}, is fully described in \citet{Sousa-2014}. Surface gravities were corrected for accuracy following \citet{Mortier-2014} and systematic errors were added in quadrature following \citet{Mortier-2018}.

\begin{table*}
  \caption{Test targets for this study. Spectral types (\textit{Sp type}) were retrieved from SIMBAD and stellar properties calculated as described in Section \ref{ss:stellar_parameters}. The apparent magnitudes in the V band (Sun: \citet{Engelke_2010}, other stars: ExoFOP-TESS) are shown in the \textit{Vmag} row. The \textit{\#Obs} row indicates the number of included spectra, while we list the median airmass for the included spectra next to \textit{Med(Airmass)}, the median SNR at 550 nm next to \textit{Med(SNR)}, the distance in time between the first and the last observation in the \textit{Time span} row, the number of observing seasons in the row denoted by \textit{\#Seasons}, and the number of clusters used to compute the cRMS in row \textit{\#Clusters}.}
  \label{tab:starproperties}
\begin{tabular}{l|rrrrrr}
        & Sun                   &  HD 127334                 & HD 62613                  &  HD 4628                  & Kepler-20                  & Kepler-21\\
\hline
Sp type           & G2                  & G5                & G9                 & K2.5               & G5                 & F6                \\
Vmag              & -26.72              & 6.36              & 6.55               & 5.74               & 12.61              & 8.25              \\
Teff {[}K{]}      & 5833  $\pm$    71   & 5758 $\pm$   76   & 5551 $\pm$   68    & 4986 $\pm$   83    & 5582 $\pm$   82    & 6400 $\pm$   94   \\
log g {[}cgs{]}   & 4.44   $\pm$   0.13 & 4.37 $\pm$   0.13 & 4.57 $\pm$   0.12  & 4.88 $\pm$   0.16  & 4.59 $\pm$   0.13  & 4.04 $\pm$   0.12 \\
Microturbulence   & 1.10   $\pm$   0.05 & 1.13 $\pm$   0.07 & 1.02 $\pm$   0.06  & 0.81 $\pm$   0.13  & 0.90  $\pm$   0.10 & 1.78 $\pm$   0.10 \\
{[}Fe/H{]}        & 0                   & 0.24 $\pm$   0.06 & -0.11 $\pm$   0.05 & -0.35 $\pm$   0.05 & 0.09 $\pm$   0.06  & 0.01 $\pm$   0.06 \\
\#Obs             & 300                 & 176             & 205                & 200                & 131                & 98                \\
Med(Airmass)      & 1.1                 & 1.07              & 1.64               & 1.19               & 1.08               & 1.19              \\
Airmass range     & 1.06 -- 2.53        & 1.03 -- 1.51      & 1.61 -- 1.99       & 1.09 -- 3.1        & 1.03 -- 1.68       & 1.02 -- 1.63      \\
Med(SNR)          & 433                 & 207               & 147                & 224                & 28                 & 177               \\
SNR range         & 367 -- 450          & 58 -- 275         & 32 -- 282          & 58 -- 363          & 13 -- 47           & 45 -- 307         \\
Time span {[}d{]} & 1003                & 405               & 628                & 405                & 1980               & 1972              \\
\#Seasons         & -                   & 2                 & 3                  & 2                  & 2                  & 4   \\
\#Clusters used         & 3                   & 3                 & 3                  & 3                  & 3                  & 3  
\\
\end{tabular}
\end{table*}

\subsection{TAPAS} \label{ss:tapas}

The Transmissions of the AtmosPhere for AStromomical data database (TAPAS; \citealt{Bertaux_2014}) takes the time and location of an observation and the position of the observed star as input and provides a transmission spectrum of the Earth's atmosphere. For now, we use only one arbitrarily chosen transmittance spectrum (La Palma Roque de los Muchachos Canarias Spain, 2018/3/30, 01:45:07, airmass 1.03)  to identify and exclude spectral regions affected by deep telluric absorption lines. More information regarding the telluric correction is provided in Section \ref{ss:tellurics}.

\section{Least-Squares Deconvolution: Background} \label{s:LSD_Background}

The Least-Squares Deconvolution (LSD) technique was proposed by \citet{Donati_1997} to measure weak magnetic signatures in noisy spectropolarimetric data. LSD has since been employed numerous times to estimate the integrated stellar longitudinal magnetic field \citep[e.g.][]{Grunhut_2013,Ramirez_Velez_2020} or to extract stellar parameters \citep[e.g.][]{Rainer_2016}. In the original application of LSD, \citet{Donati_1997} showed that in the weak-field regime for weak lines, a Stokes V spectrum can be modelled via the convolution $M_{\text{c}}\ast \text{Z}$, where $M_{\text{c}}$ is a sum of delta functions centred at the absorption lines' wavelength weighted by the individual lines' central depths while $\text{Z}$ is the LSD common profile to which all weak lines are similar. In this work, we use the same technique to model intensity (Stokes I) spectra. A Stokes I RV extraction via LSD is also mentioned in e.g. \citet{Barnes_2012} and \citet{Heitzmann_2021}. In \citet{Barnes_2012}, the LSD technique was applied on M dwarf spectra obtained with the Magellan Inamori Kyocera Echelle Spectrograph \citep{Bernstein_2003} at R $\sim$ 35,000. Utilising the 620--900 nm region, they reached 10\ms precision RVs using telluric absorption lines as reference fiducial. \citet{Barnes_2012} cross-correlated against the telluric lines due to the lack of a sufficiently precise wavelength solution, but relying on the tellurics as wavelength reference can only provide limited accuracy due to unkown wind speeds in the Earth's atmosphere along the line of sight. 
In this study, we analyse spectra with a high-fidelity wavelength solution obtained with pressure and temperature stabilised spectrographs. Also, in contrast to \citet{Heitzmann_2021}, we focus specifically on LSD as a tool to extract high-precision RVs from high-resolution intensity spectra.

\subsection{Stokes I LSD} \label{ss:stokes_i_lsd}

By nature of the convolution, the LSD technique relies on the assumption that the profiles of overlapping absorption lines add up linearly. This assumption is valid for absorption lines blended due to e.g. rotational broadening, but not for the blending caused by intrinsic overlap of the lines \citep[e.g.][]{Kochukhov_2010}. 
Because the spectra of M-dwarfs are dominated by blended molecular absorption lines, LSD RV extraction is not well suited for those stars.
A-type and earlier stars are unsuitable for precision-RV studies in general because of their high rotation rates \citep[e.g. of order 100--200 \kms\ for A-type stars,][]{Galland_2005}, and the relative scarcity of absorption lines due to the higher surface temperature. 
We therefore apply the LSD method only to main sequence FGK-type stars in this study.

To mitigate model insufficiencies, we exclude wavelength regions where the convolution does not match the spectrum very well, as explained in Section \ref{ss:abslineselection}. This excludes blended absorption lines for which the assumption of linear addition is clearly not met.

As outlined above, we assume that the spectrum can be modelled by the convolution product $I = M_{\text{c}}\ast Z$, where the LSD common profile $Z$ is to be determined and carries the RV information. The method described in the following paragraphs is also explained in \citet{Donati_1997}. To extract $Z$, we first define the following quantities:
\begin{itemize}
    \item $\lambda_i$: wavelength of pixel i.
    \item $\lambda_l$: central rest frame wavelength of absorption line $l$.
    \item $w_l$: weight of absorption line $l$, set to its central depth.
    \item $v_{il} = c \frac{\lambda_i - \lambda_l }{\lambda_l}$: radial velocity $v_{il}$ which shifts $\lambda_l$ to $\lambda_i$.
    \item $v_j$: velocity grid point $j$ of LSD common profile $\text{Z}$.
\end{itemize}
With these definitions, the convolution product $I = M_{\text{c}}\ast Z$ can be expressed as the matrix multiplication $\text{I} = \textbf{M} \cdot \text{Z}$ using
\begin{equation}
    \textbf{M}_{ij} = \sum_{l} w_l \: \Lambda\left(\frac{ v_j - v_{il} }{\Delta v}  \right),
\end{equation}
where $l$ is the index of the absorption line in the line list, $i$ stands for the pixel number, $j$ is the index of the respective LSD common profile point, $\Delta v$ is the velocity increment between two adjacent points of the LSD common profile, and the function $\Lambda$ is defined as
\begin{equation}
    \Lambda \left( x \right) = \begin{cases}
    0 & \left|x\right| \geq 1 \\
    1 - \left|x\right| & \left|x\right| < 1.
    \end{cases}
\end{equation}
If $v_{il}$ is exactly equal to $v_j$, then the central wavelength of absorption line $l$ in the rest frame and the wavelength associated with pixel $i$ are $v_j$ apart in velocity space and thus $\text{Z}_j$ is added to the LSD model for pixel $i$. If the value of $v_{il}$ is between two velocity grid values of $\text{Z}$, we interpolate linearly and add a fraction of $\text{Z}_j$ to the LSD model for pixel $i$.

To find the best-fitting LSD model to the data, we minimise
\begin{equation} \label{eq:minimisation}
\chi ^2 =  (\text{Y} - \textbf{M} \text{Z})^\intercal \textbf{S} (\text{Y} - \textbf{M} \text{Z}),
\end{equation}
where Y is the deblazed and normalised spectrum, and $\textbf{S}$ is a diagonal matrix with the diagonal entry in row $i$ containing the inverse square of the uncertainty of $Y_i$.
By applying some linear algebra (cf. Appendix \ref{app:Zderivation}), an analytic expression for the common profile $\text{Z}$ minimising the residuals $\chi$ can be found:
\begin{equation} \label{eq:deconvolution}
\text{Z} =   (\textbf{M}^\intercal \textbf{S} \textbf{M})^{-1}\textbf{M}^\intercal \textbf{S} \text{Y}.
\end{equation}
In this work, we use the described framework on the individual orders of the Stokes I spectra setting the weight $w_l$ to the absorption line central depth $d_l$ from VALD3. The diagonal entries of $(\textbf{M}^\intercal \textbf{S} \textbf{M})^{-1}$ represent the variances of the resulting common profile as a standard result of weighted least-squares fitting \citep[cf.][]{Hogg_2010}.

To get a high-SNR master common profile, we compute a weighted sum of the order common profiles (cf. Section \ref{ss:order_weighting}) and we propagate the uncertainties of the individual order common profiles, as calculated above, to the master LSD common profile.

Lastly, we fit a Gaussian to the master LSD common profile, using the propagated inverse associated variances as weights, to extract the RV estimate. For the stars analysed in this study, fitting a Voigt profile did not yield lower scatter time series as compared to fitting a Gaussian. Since the pseudo-continuum and the outer wings can be noisy, we progressively cut the outer parts of the LSD common profile at steps of a quarter half width at half maximum (HWHM) of the first LSD common profile until the time series RV Root Mean Square (RMS) reaches a minimum. The same cut is used for all spectra for the same star. For the stars in this study, the cut never removes LSD common profile points within the initial HWHM.
The uncertainty associated with an RV estimate can be computed from the common profile uncertainties following eq A.2 and A.3 in \citet{Boisse_2010}:
\begin{equation}
    \sigma_{\text{RV}} = \left(   \sum_j \left(\frac{\partial CP(j)}{\partial v(j)}\right)^2  \frac{1}{\sigma_{CP(j)}^2} N_{\text{scale}}    \right)^{-0.5},
\end{equation}
where $CP$ is the LSD common profile, $\sigma_{CP}$ its uncertainty, $v$ the velocity grid, and $N_{\text{scale}}$ the velocity step in detector pixel units, which is 1 in our case (cf. Section \ref{ss:velocity_grid}).

\section{Stellar Continuum Correction} \label{s:Continuum_correction}

Stellar absorption lines are embedded in the stellar continuum resembling the black-body radiation spectrum at the star's effective temperature. For spectra measured with ground-based instruments, the shape of the continuum is altered due to atmospheric extinction reducing the photon flux, most notably at shorter wavelengths. We absorb the smooth, slowly-varying component of atmospheric extinction into the continuum estimate since this is equivalent to treating them separately.

A continuum estimate is necessary because the non-flat continuum distorts the shape of the absorption lines and we need the absorption lines to be normalised to model them using the relative line depth from VALD3. Thus, we divide each spectrum by its corresponding continuum, as done in all RV pipelines that we have investigated. Since an individual pixel measures photons with wavelength between $\lambda_1$ and $\lambda_2$ ($\lambda_2>\lambda_1$), a background function that is increasing towards longer wavelengths (blaze and continuum) leads to the pixel measuring more photons closer to $\lambda_2$ than to $\lambda_1$. This overweighs the part of the stellar spectrum closer to $\lambda_2$. Dividing by a binned continuum does not take this into account. Our simulations show this approximately results in a residual RV error per absorption line of about 10\cms for HARPS-N depending on the local curvature of the continuum and the blaze function, and the wavelength range measured by a pixel. We leave the further investigation or correction of this effect to future work. 

A multitude of different continuum estimation techniques has been applied in the literature \citep[e.g.][]{Rainer_2016,Dumusque_2018,Bedell_2019}. While most techniques rely on iteratively masking absorption lines and fitting a polynomial to the remaining parts of the spectrum, the RASSINE code \citep{Cretignier_2020}, which we employ in this analysis, uses alpha shape algorithms to find the upper envelope of a spectrum spanned by the local maxima in the spectrum. In the two-dimensional case, the alpha shape can be thought of as rolling a disc of radius $\alpha$ over the given points and connecting each two points touched by the disc. Smaller values of $\alpha$ reveal more cavities \citep{Edelsbrunner_1994}. The assumption that this envelope corresponds to the continuum is well satisfied for solar-like stars, except where many absorption lines overlap, such as in the bluest part of the visible range \citep{Cretignier_2020}. This is not worrisome in our case because most RV information is extracted from the higher echelle orders as the signal-to-noise is low in the very blue (i.e. from 383 nm up to about 410 nm, cf. Section \ref{ss:weight_matrix}), and many absorption lines in this wavelength region are masked due to high model deviations (cf. Section \ref{sss:dd_lineselection}). Moreover, \citet{Cretignier_2020} showed that this method still provides more precise results compared to the classical continuum-normalisation techniques.

RASSINE is meant to be used on merged blaze-corrected echelle order spectra (1D/S1D). Since we intend to continuum-normalise unmerged (2D/S2D) spectra, an extra step is needed. Applying RASSINE on every single order is not a viable option since this yields dissatisfactory results if there are wide absorption lines present. 

\subsection{2D RASSINE} \label{sss:2drassine}

By emulating some of the steps taken to generate the S1D spectra, the echelle order spectra can be brought onto the same flux level as the S1D spectra. The S2D spectra can then be normalised with the RASSINE continuum computed for the S1D spectra by interpolating the continuum to the wavelength solution of the S2D spectra. By this procedure, we avoid interpolating the echelle order spectra and any distortion to the absorption lines which can result from this step, but we can still use RASSINE as intended by the authors without any alterations to the code. For now, we assume that the blaze-corrected fluxes of overlapping wavelength parts do not differ by much. This assumption is justified and analysed in Section \ref{sss:fluxdiscrepancy}.

To create an S1D spectrum, the DRS (version 2.2.2 and 2.3.1) reads in the S2D spectra and stitches the 69 blaze-corrected echelle order spectra together. Thereafter, it reinterpolates the fluxes on a grid with constant velocity increment (0.82\kms). During the interpolation process, the DRS pipeline indirectly takes into account that each pixel sees a different wavelength range. The wavelength grid is then shifted to the barycentric frame of reference. To emulate these steps, we first read these data:

\begin{itemize}
\item Deblazed S2D spectra (spectra extracted from CCD divided by blaze function).
\item Wavelengths $\lambda$ in barycentric frame of reference.
\item Wavelength steps between two adjacent pixels $d_{\lambda_i}$.
\end{itemize}

The deblazed pixel fluxes are multiplied by $f_{\text{corr}}$ to linearly scale the pixel fluxes from the wavelength range that a pixel measures to the wavelength step used in the S1D spectrum.
More specifically, this correction factor $f_{\text{corr}}$ is equal to $\frac{0.82\, \lambda_i}{c\, d_{\lambda_i}}$, where c is the speed of light in \kms and $d_{\lambda_i}$ is the wavelength step between two pixels in units of \AA. This factor arises due to the wavelength range associated with a given S1D data point being equal to $\frac{0.82\, \lambda_i}{c}$. \footnote{The earlier version of the DRS (see https://eso.org/pub/dfs/pipelines/instruments/espresso/espdr-pipeline-manual-2.3.3.pdf) reinterpolates the fluxes on a grid with a constant wavelength increment of 0.01 \AA. The correction factor $f_{\text{corr}}$ would then simply be $\frac{0.01}{d_{\lambda_i}}$, with $d_{\lambda_i}$ being again the wavelength step between two pixels in units of \AA.}

\begin{figure}
	\includegraphics[width=\columnwidth]{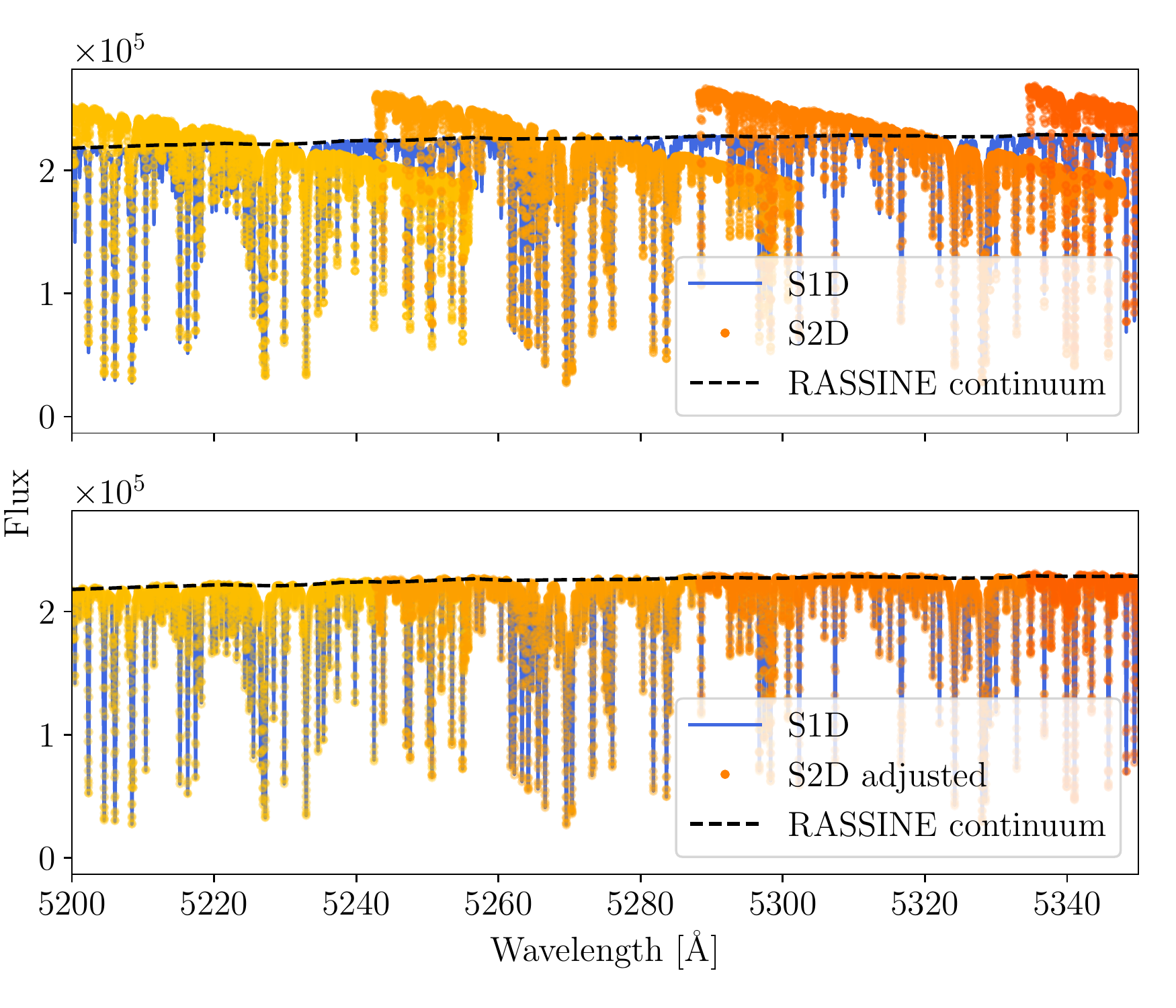}
    \caption{In the upper panel, the deblazed spectra of four echelle orders are displayed on top of the corresponding S1D spectrum. In the lower panel, the S2D echelle spectra have been adjusted, as described in Section \ref{sss:2drassine}, to match the S1D spectrum and, consequently, the RASSINE continuum.}
    \label{fig:rassine2d1d}
\end{figure}

Some original deblazed S2D orders and their wavelength step adjusted version are displayed in Fig. \ref{fig:rassine2d1d}. For now, this procedure works for any spectrograph producing S1D spectra in the same fashion as HARPS. If 1D merged spectra were not be available, they could be straightforwardly produced following the above procedures.

\subsection{Overlap Discrepancy} \label{sss:fluxdiscrepancy}

As mentioned in Section \ref{sss:2drassine}, we assume that the RASSINE continuum computed for the S1D spectrum is suitable for the individual orders if these are adjusted to match the S1D flux level. Since adjacent spectral orders overlap in wavelength, these parts of the spectrum are essentially measured with two different sets of pixels. To create an S1D spectrum, the weighted average of the overlapping parts is computed after deblazing, which is what we called ``stitching'' in Section \ref{sss:2drassine}. If the flux for the same wavelength noticeably differs between two orders, the S1D flux will not match either of them. Consequently, the RASSINE continuum will not be perfectly suitable for the overlapping part of these orders.

To quantify the discrepancy between the fluxes of overlapping orders, we compute
\begin{equation}
    \Delta = \text{med}_{\text{orders}} \left( \text{med}_{\lambda} \left(\mid f_\text{o}(\lambda) - f_\text{o+1}(\lambda) \mid \right)\right),
\end{equation}
which is the median over all orders of the median absolute difference between the fluxes within the overlapping region. The discrepancy $\Delta$ is always greater than zero since we compute the absolute value of the flux differences to avoid cancellation effects.

\begin{figure}
	\includegraphics[width=\columnwidth]{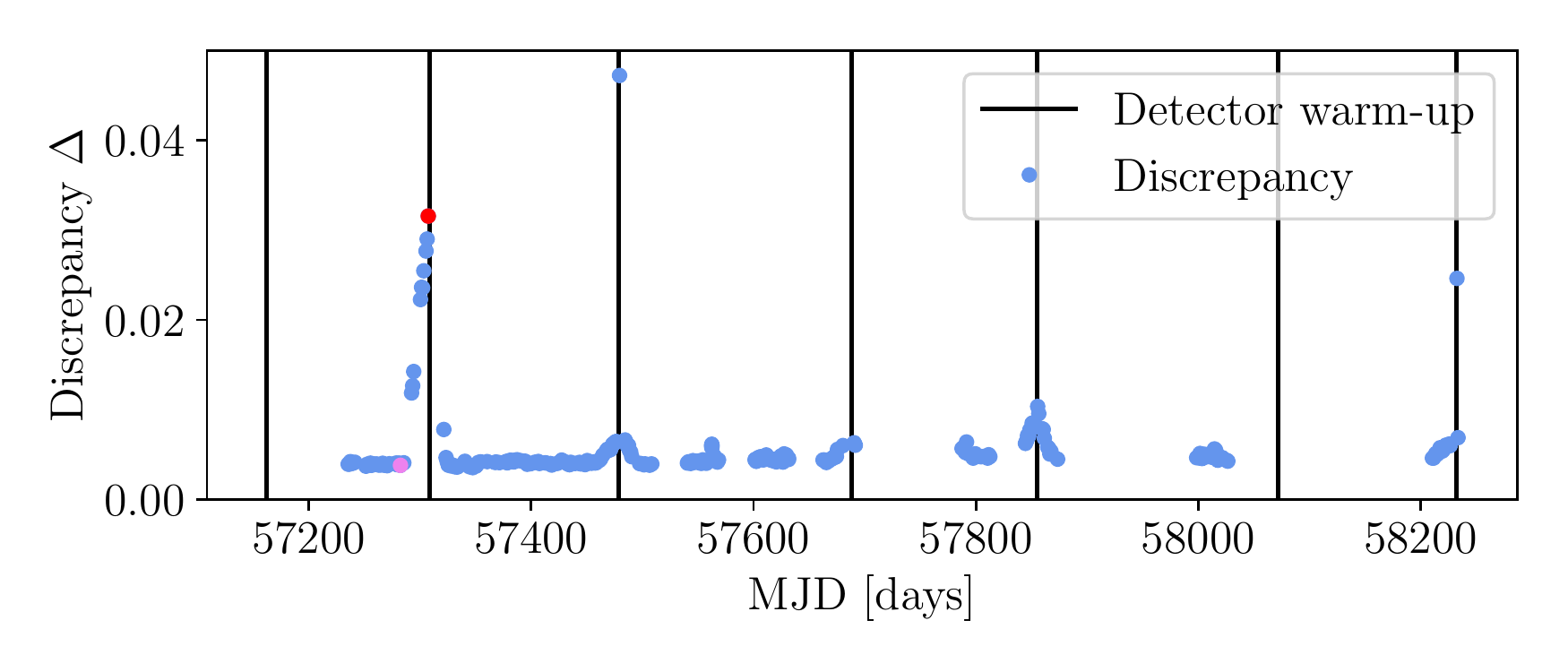}
    \caption{Discrepancy plot for 300 spectra in three years of HARPS-N solar observations. The discrepancy between the fluxes of overlapping orders is typically very low. As the humidity increases, the discrepancy increases. After a detector warm-up, the discrepancy reduces rapidly and settles to a low level. An example for a low (high) discrepancy spectrum is marked in violet (red) and shown in Fig. \ref{fig:fluxdisc_correction}.}
    \label{fig:fluxdisc_timeseries}
\end{figure}

\begin{figure}
	\includegraphics[width=\columnwidth]{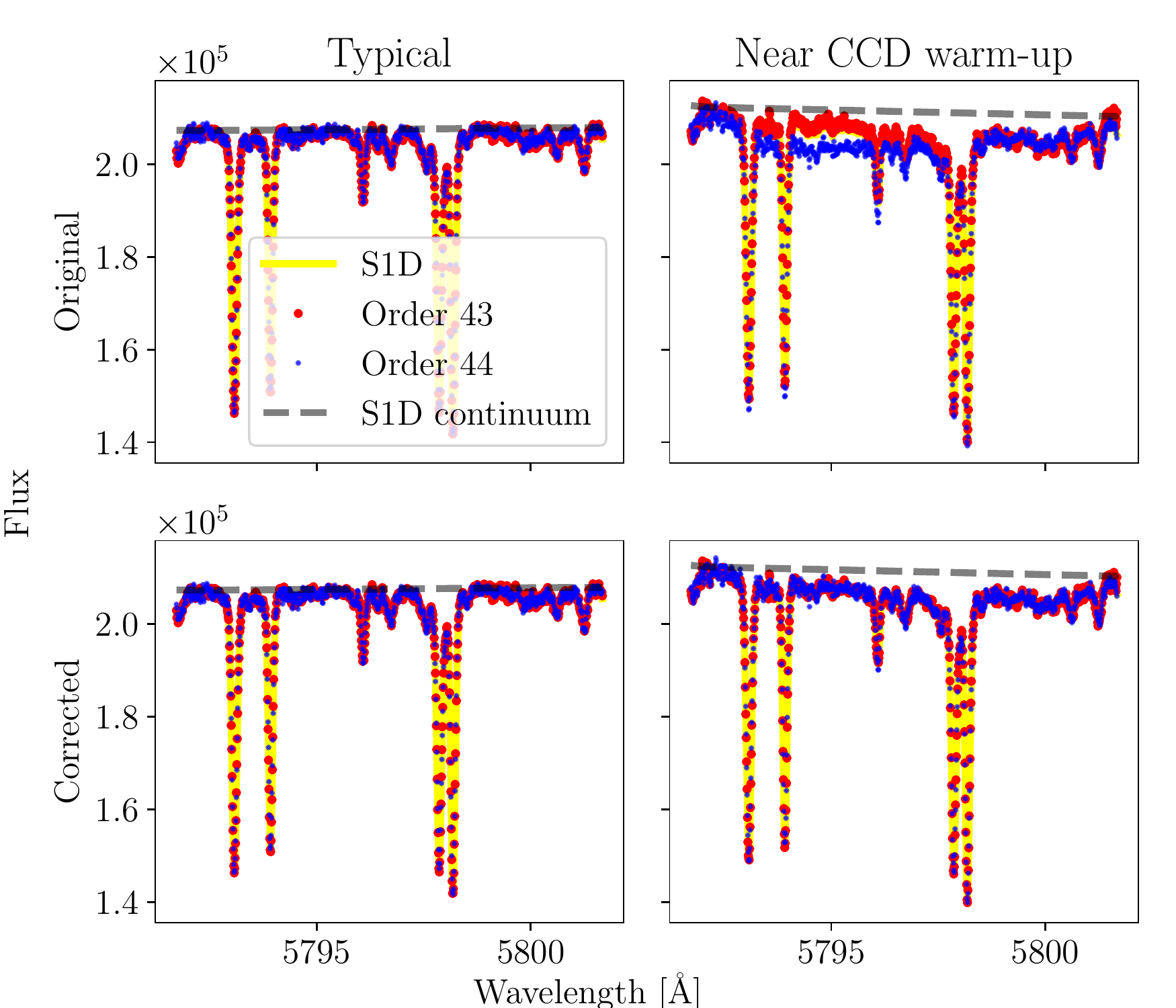}
    \caption{The upper two panels show the deblazed fluxes of two adjacent orders where they overlap in wavelength. On the left, we display a typical spectrum for which there is no overlap flux discrepancy (marked in violet in Fig. \ref{fig:fluxdisc_timeseries}). On the right, a spectrum just before a CCD warm-up is displayed (marked in red in Fig. \ref{fig:fluxdisc_timeseries}). The fluxes of different pixels measuring the same wavelength region visibly do no match in this case. In the bottom panels, we show the same spectra but adjusted such that both are on the same level as the S1D spectrum.}
    \label{fig:fluxdisc_correction}
\end{figure}

Until October 2021, the HARPS-N spectrograph had a small leak in the cryostat \citep{Dumusque_2021}. This meant that over time the humidity increased, subsequently increasing the reflectivity of the detector. This issue was resolved by warming up the detector roughly every 6 months. As visible in Fig. \ref{fig:fluxdisc_timeseries}, the overlap flux discrepancy increased towards these detector warm-ups as the humidity and the reflectivity of the detector increased. Most spectra were not significantly affected by this effect and resemble the spectrum on the left in Fig. \ref{fig:fluxdisc_correction}. 

To correct for the larger discrepancies, we divide the overlapping part of each order by the smoothed ratio between its flux values and the S1D flux values. This brings the fluxes of both orders onto the flux level of the S1D spectrum, as depicted in the bottom panels of Fig. \ref{fig:fluxdisc_correction}. In other words, the continua of both orders now lie on top of the s1d continuum and thus we can use the RASSINE continuum to normalise these spectra. The RVs of the individual spectra are only marginally affected by this procedure. Near the detector warm-ups, the overlap correction altered the solar RVs by less than 10\cms and typically overall by less than 1\cms.

\section{Core LSD Algorithm} \label{s:LSD_algo_description}

In this section, we describe the application of LSD to our normalised ordered spectra and the steps that are taken to exclude data that adversely impact the RVs. There are several effects other than stellar activity altering the estimated Doppler shift of a spectrum or individual absorption line. For instance, a curved continuum or the overlap with a neighbouring absorption line wing can shift the RV of a single line. Other effects include telluric lines and insufficiencies in the model which are incorrectly compensated for in the least-squares minimisation by altering the shape of the LSD common profile. To mitigate this effect, we may remove the data that is likely to be affected. However, excluding problematic wavelength regions too vigorously also increases the RV scatter. This is due to the reduction in the number of included absorption lines, which discard the RV information in distorted lines. Including distorted lines can be beneficial since the RV errors cancel out across the spectrum. Excluding data thus has to be performed carefully. In this section, we identify several effects and denote the associated parameters using bold Greek letters (see list in Section \ref{ss:findparameters}).

\subsection{Absorption Line Selection} \label{ss:abslineselection}

\subsubsection{Data driven line selection} \label{sss:dd_lineselection}
\begin{figure}
	\includegraphics[width=\columnwidth]{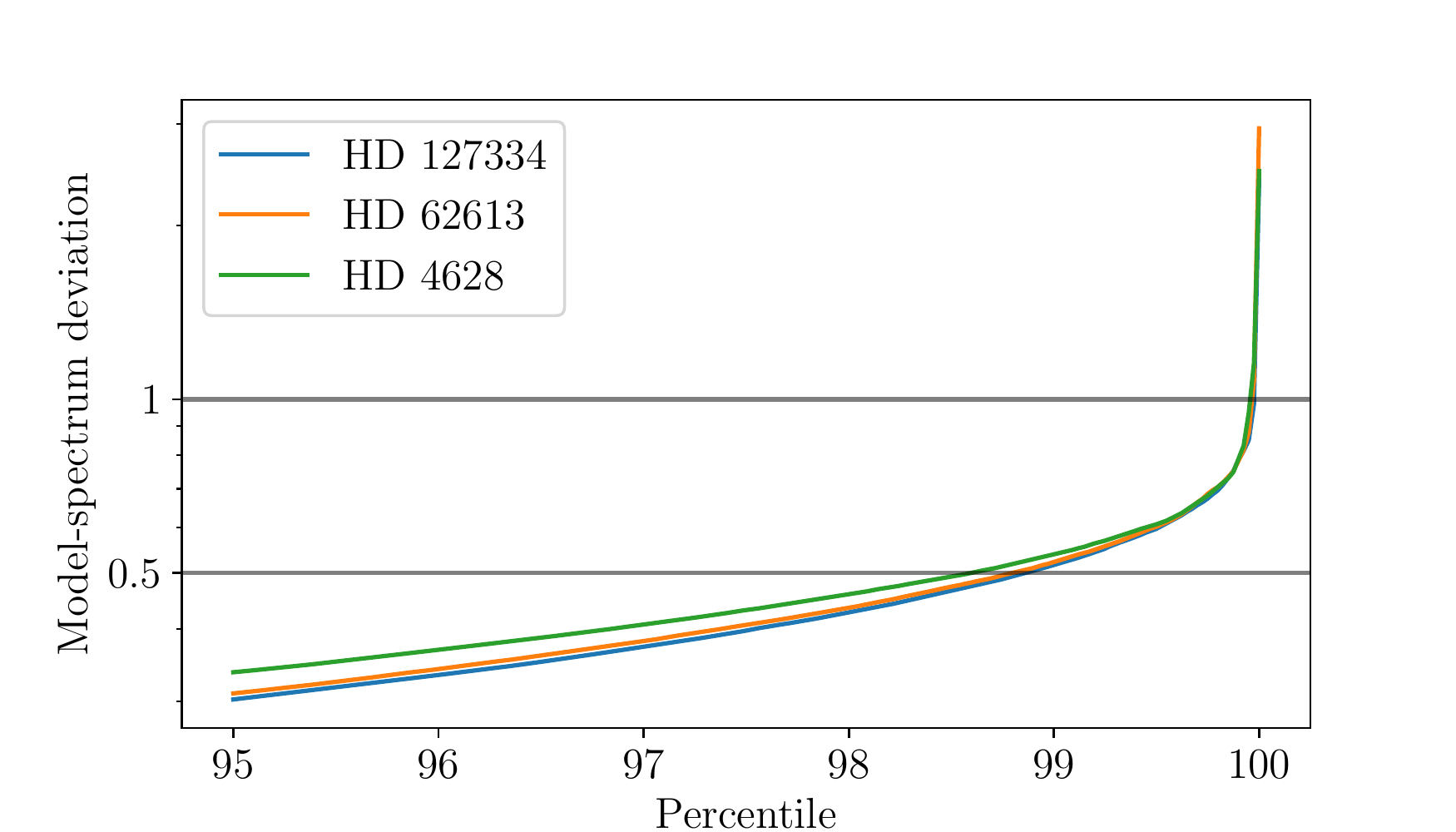}
    \caption{Distribution of the absolute deviation between the first convolution model and the spectrum for three stars with no known planets. The horizontal lines indicate where the absolute difference between the model and the spectrum is about 0.5 and 1.}
    \label{fig:ms_deviation}
\end{figure}

To get a first LSD model, we run the LSD code on the first spectrum of the time series without masking any data points except telluric lines deeper than 10 per cent relative to the continuum. The convolution of the resulting LSD common profile with the line list provides us with a first model of the spectrum. This model can be directly compared to the spectrum to identify spectral regions where the LSD assumptions are not valid or where there is a discrepancy between the estimated and the actual depth of the absorption lines.
To mitigate the impact of these lines, we flag all pixels for which the absolute deviation between the model and the spectrum is greater than $\mathbf{\Gamma}$. Any flux within half the velocity grid width of a flagged pixel is then excluded. The effect is star dependent and thus a suitable $\mathbf{\Gamma}$ is different for each star, but it is typically about 0.5. The percentage of fluxes with model-spectrum deviation below a given value are shown in Fig. \ref{fig:ms_deviation}.
About 99 per cent of the normalised fluxes differ by less than 0.5 from the convolution model, while the model-spectrum deviation is below 1 for about 99.95 per cent of the fluxes.

\subsubsection{A priori line selection} \label{sss:ap_lineselection}

For this analysis, we include all VALD3 absorption lines deeper than 0.01 (1 per cent relative to the normalised continuum, see Section \ref{ss:vald}) for the deconvolution, but we only include spectral regions that contain a line with relative depth deeper than 0.1 and no line deeper than $\mathbf{\Xi}$. The first threshold can be varied in the code, but no consistent significant RV improvement for lines below the accepted depth range of 0.1 was noticed. Neither did we find a consistent improvement from only including lines deeper than 0.2 which is the global threshold in \citet{Heitzmann_2021}. Excluding spectral regions with no prominent absorption lines is beneficial since the RV information content in weak absorption lines is marginal compared to deeper lines and weak lines are more easily perturbed by telluric lines, instrumental systematics, and stellar activity \citep[e.g.][]{Cretignier_2020_2}.

By setting an upper depth threshold $\mathbf{\Xi}$, we exclude all fluxes with associated wavelength closer than half the width of the velocity grid window (i.e. half the velocity span covered by the common profile) to a line deeper than $\mathbf{\Xi}$.
The optimal value for this threshold depends primarily on whether a good fit to the absorption lines or a precise RV is preferred.
\citet{Kochukhov_2010} found that the line similarity assumption intrinsic to LSD is only well satisfied for spectral lines weaker than 0.4 in intensity spectra. 
A $\mathbf{\Xi}$ in this regime is thus expected to be beneficial for analyses that rely on a good fit of the convolution model to the spectrum.
In this study, we set a high upper threshold $\mathbf{\Xi}$ to include the RV information in the deep lines and extract precise RVs. To avoid RV scatter due to a lack of line similarity or insufficiencies in the model, we vary the maximum depth $\mathbf{\Xi}$ to exclude the very deepest lines if they induce additional scatter as outlined in Section \ref{sss:multimaskapproach}.

Another potential issue is that wide absorption wings distort the shape of the adjacent absorption lines. These distorted lines cannot be properly modeled by a single convolution and thus introduce RV errors. We tested identifying lines with wide wings \textit{a priori} from the VALD3 parameters and also fitting Gaussian profiles to each single absorption line to identify the deep wider lines such as H-alpha or the Sodium D lines. Excluding these lines only marginally changed the RV scatter (less than 5\cms) and is thus not included by default. Since these wide lines can distort the shape of the common profile, which might be used for other purposes than RV extraction, an option is provided to exclude wide lines.


The same wavelengths in the barycentric reference frame are excluded for the other spectra of the same star.

\subsection{Velocity Grid Window Width} \label{ss:velocity_grid}

We only include data that is within reach of the line list convolved with the LSD common profile, i.e. fluxes that are less than half the velocity window width apart from at least one absorption line in the VALD3 line list. While the velocity step of the velocity grid is set to match the typical wavelength step between two pixels, i.e. 0.82 km/s, and the central velocity point can be set to the radial velocity of the star in the first spectrum, determining the window width of the velocity grid on which the LSD common profile is computed is less straightforward. The width of the velocity grid window is set to $\mathbf{\Phi}$ times the FWHM from the first LSD run conducted as described in Section \ref{sss:dd_lineselection}.

\subsection{Telluric Absorption Line Correction and Masking} \label{ss:tellurics}

The Earth's atmosphere's absorption lines (telluric lines) are superimposed on the stellar absorption lines. While the stellar absorption lines shift in wavelength as measured in the observatory rest frame, the telluric absorption lines practically remain at the same wavelengths. Telluric lines vary in width and depth, however. \citet{Cunha_2014} showed that even micro-tellurics can induce RV errors of up to 1\ms in HARPS spectra using the CCF technique.\\
In this code, we take one telluric transmission spectrum from TAPAS with relatively weak telluric lines, which means that dividing by this model will reduce the depth of the tellurics in the spectrum without putting the corrected fluxes above the continuum level. Using the same transmittance spectrum for all spectra is clearly not ideal. The code is set up to ingest (and normalise if necessary) a user-provided telluric transmittance model as a matrix or in the TAPAS fits format. 

Since the step above is not sufficient, we additionally remove wavelengths affected by deep tellurics. More specifically, we first flag all wavelengths with telluric line depth above $\mathbf{\Theta}$ (\textit{telluric map v1}). This mask is then shifted to the barycentric frame of reference for each spectrum. From these masks, we create a master telluric mask (\textit{telluric map v2}) which flags any flux value if its corresponding barycentric wavelength was masked in any of the spectra. This means that the mask is essentially fixed in the barycentric frame of reference and the same stellar and telluric absorption lines are excluded for all spectra. This is to ensure that the same stellar absorption lines are included for all spectra while consistently excluding all absorption lines affected by deep telluric lines.
It is important to always include the same absorption lines because the theoretical absorption line wavelengths' uncertainties are of the order of 100 \ms. This translates to varying RV shifts if lines are not consistently included or excluded. 

\begin{figure}
	\includegraphics[width=\columnwidth]{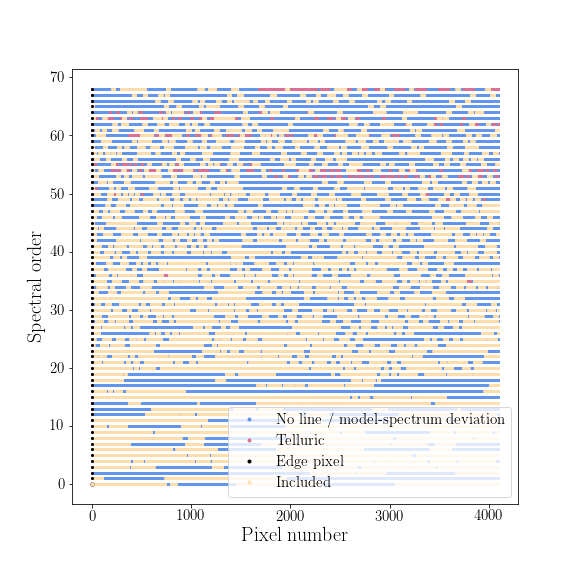}
    \caption{Pixel quality map for all orders of the S2D spectra, showing which pixels are excluded or included after the various steps. Only the pale orange pixels are included for this spectrum.}
    \label{fig:inclusion_map}
\end{figure}

An example for the excluded and included pixels is shown in Fig. \ref{fig:inclusion_map}. The pixels marked in blue in Fig. \ref{fig:inclusion_map} are excluded for the given spectrum because there is either no absorption line present or the deviation between the spectrum and the convolution model is too high. 

Absorption lines are shifted due to the Earths orbital motion which can carry absorption lines off the edge of an order for some spectra. The pixels for which this is the case are marked in black and excluded for all spectra, as in \citet{Donati_1997}, since including the respective absorption lines can lead to systematics, as outlined above. The pixels marked in red are affected by telluric absorption lines and thus excluded too. Note that we consistently exclude the same wavelengths in the barycentric reference frame for all spectra and consequently the pixel mask shifts slightly with the observatory's motion relative to the solar system's barycentre.

\subsection{Weight Matrix} \label{ss:weight_matrix}

\begin{figure}
	\includegraphics[width=\columnwidth]{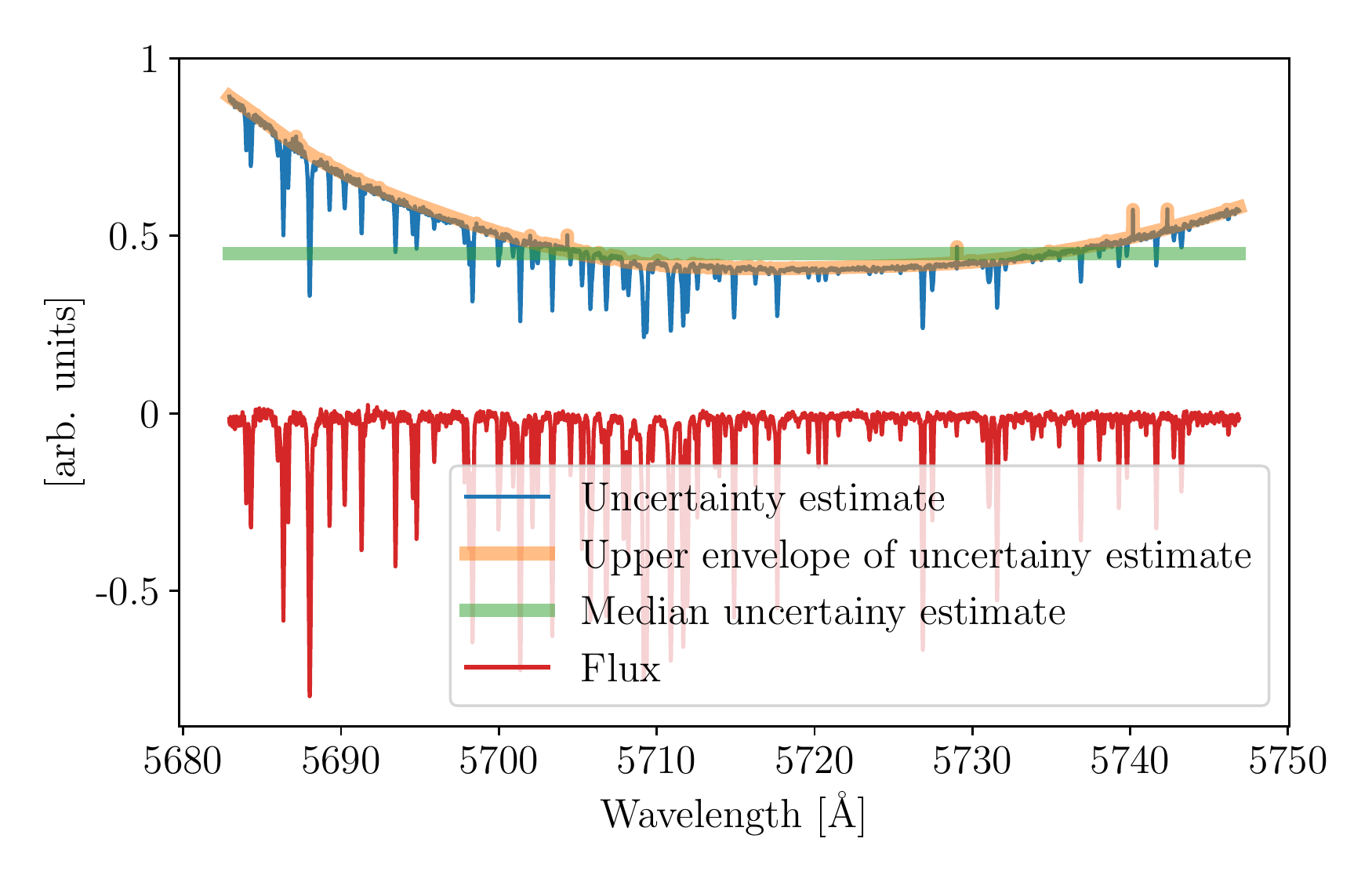}
    \caption{Uncertainty estimate of one echelle order for HD 4628. The uncertainties are scaled by a factor 100 and the fluxes are normalised and shifted to 0. The curved shape of the uncertainty is mainly caused by the blaze function.}
    \label{fig:upperenvelope}
\end{figure}

The weight matrix $\textbf{S}$ (cf. Eq. \ref{eq:minimisation}) is used to account for the uncertainty of the flux values in the deconvolution process. Essentially, in the least-squares minimisation, the deviation of the LSD convolution model from the flux values is weighted by the respective diagonal entry in the weight matrix. For these weights, we tested the inverse square of the flux uncertainty estimate (hereafter \textbf{flux weights}), the inverse square of the upper envelope of the flux uncertainty estimate except where there are uncertainty outliers due to e.g. cosmic rays (\textbf{envelope weights}), and the inverse square of the median flux uncertainty estimate (\textbf{uniform weights}). The underlying error estimates for one order are displayed in Fig. \ref{fig:upperenvelope}. There are three interconnected reasons for testing weighting schemes that differ from the standard flux weighting scheme:

Firstly, using the flux uncertainties assigns a high weight to the deeper lines while almost ignoring the RV information in the weaker lines. This can lead to problems if these highly-weighted absorption lines are contaminated by tellurics, stellar activity, or instrumental problems.

Secondly, each absorption line has its own impact on the LSD common profile shape and shift. As there is necessarily a slight discrepancy between the true and the listed absorption line wavelength, the RV measured from a single line has a slight RV offset (cf. Section \ref{ss:tellurics}). This is negligible if the weight, and thus the RV contribution of an individual line, is constant throughout the time series. An individual absorption line's weight varies in time, however, due to the varying airmass and the Earth's motion. The induced RV errors cancel out better if there are no lines with very high weight. This RV effect due to the varying weight of an individual line is most pronounced near the edges of an order, where the gradient of the blaze function is highest. 

Thirdly, a simple analysis shows that the centroid of a Gaussian absorption line with a minor other absorption line at slightly different wavelength is better fit by one Gaussian with constant weight applied rather than one Gaussian with the inverse photon errors as weights. Since there are stellar absorption lines missing in the line list, unaccounted micro-tellurics in the telluric model, and the convolution-intrinsic line addition can lead to problems for blended (and thus often deeper) lines, setting the weights to the inverse squared uncertainty is not optimal. 

For the three reasons listed above, we tested the inverse square of the envelope of the flux uncertainty estimate. This weighting scheme is supposed to avoid overweighing potentially variable deep lines, while also giving less weight to more variable parts of the spectrum near the order edges.
Furthermore, we tested uniform weights by setting the uncertainty of all fluxes within an order to the median uncertainty of the given order.
Using uniform weights per order has the advantage that the weight of an absorption line is independent of the observatory's motion relative to the star, which is desirable as explained above. However, noisy data points at the edge of an echelle order are weighted the same as high-quality data in the centre of the same order, which can adversely impact the RV estimate.

We also tested updating the uncertainty estimate based on the flux variance of a wavelength bin in the stellar frame of reference adjusted for the photon noise. This procedure did not decrease the scatter in the considered RV time series and is thus not included in this study.

The final results for all three weighting schemes are shown in Table \ref{impr_per_step_stars}.

\subsection{Order weighting} \label{ss:order_weighting}

\begin{figure}
	\includegraphics[width=\columnwidth]{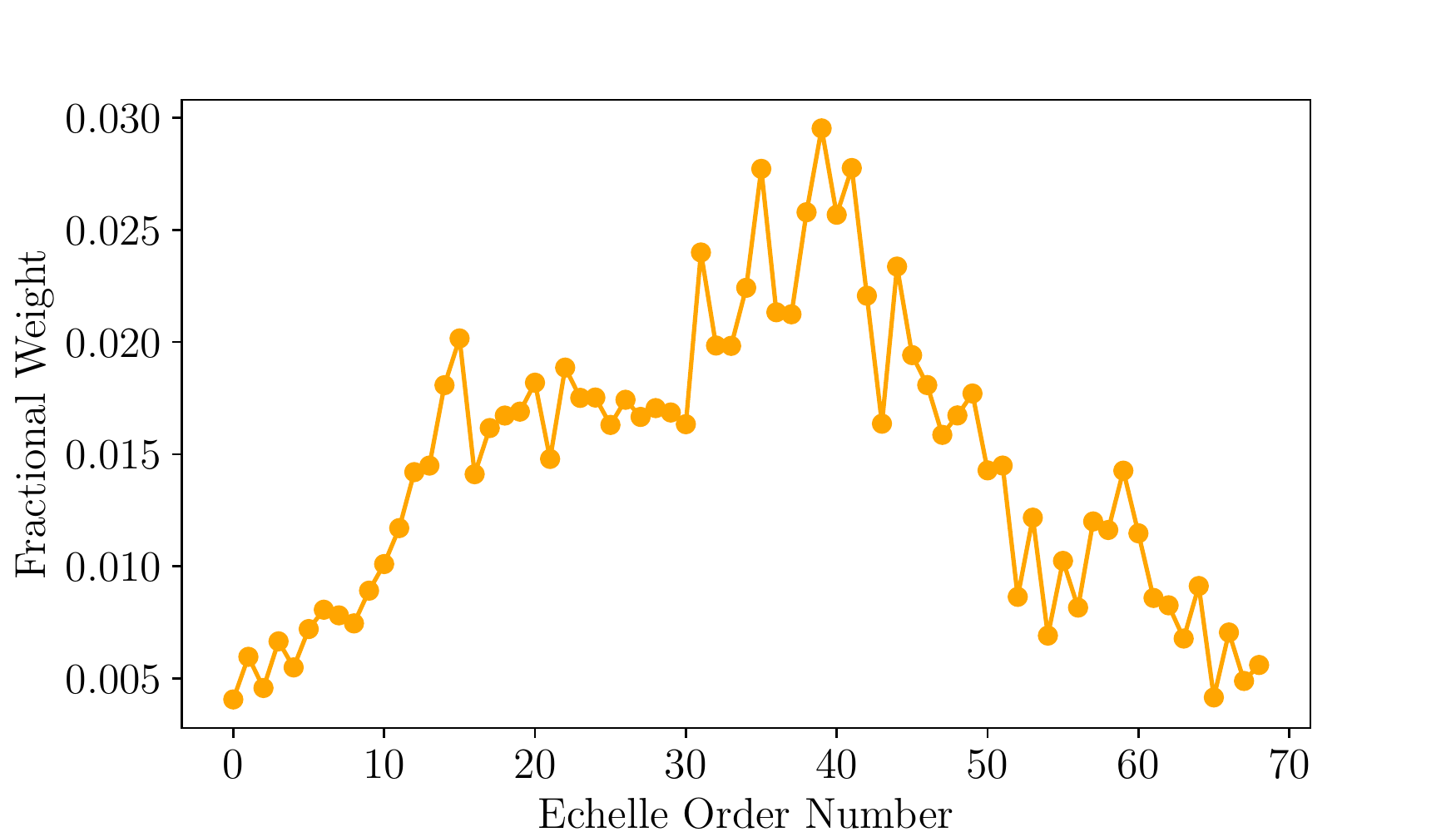}
    \caption{Fractional weight of an individual spectrum. The master LSD common profiles are produced by adding the common profiles of the individual orders multiplied by the order weight shown in this figure.}
    \label{fig:order_weight}
\end{figure}

After running the deconvolution on each order, we are left with 69 common profiles. To combine these to one master LSD common profile with higher signal-to-noise, we compute the weighted mean of the individual LSD common profiles. Suitable weights can be computed by summing the weights of all included pixels within a given order given by the entries in the weight matrix as described in Section \ref{ss:weight_matrix}. Orders with only few absorption lines or only very noisy flux values therefore get a comparatively lower weight. By default, this is executed for all spectra in the time series. A typical example of the order weight can be seen in Fig. \ref{fig:order_weight}. To assess the variability of the order weights in time, we computed the standard deviation of the individual normalised order weights. Unsurprisingly, the weights of orders with low weight are more variable than the high-weighted orders since the low-weighted orders include fewer pixels (cf. Fig. \ref{fig:inclusion_map}) and more variable parts of the spectrum. The weights of the highest-weighted orders vary by about 3 per cent, while the median standard deviation is typically about 5 to 10 per cent.
If deemed advantageous, the weighting scheme can easily be changed to keep the weights constant throughout the time series. We noticed no significant difference in RV scatter in our test cases. Note that the order weights are also used to propagate the uncertainty estimates of the order common profiles to the master LSD common profile.

\section{Parameter optimisation}  \label{ss:findparameters}

The RV scatter of the resulting time series depends on how the parameters listed below are set:

\begin{itemize}
\item[$\mathbf{\Gamma}$]
Absolute difference between spectrum and convolution model ($\text{Y}$-$\mathbf{M}\cdot Z$).

\item[$\mathbf{\Xi}$]
Maximum depth of included VALD3 absorption lines.

\item[$\mathbf{\Phi}$] 
Velocity window size over which LSD is carried out, in units of LSD line FWHM.

\item[$\mathbf{\Theta}$]
Depth threshold for the telluric mask. Any part of the stellar spectrum that is ever touched by a telluric line deeper than $\mathbf{\Theta}$ is excluded for the entire time series. A depth threshold $\mathbf{\Theta}$ of 0.1 is equivalent to excluding telluric lines with transmission below 90 per cent.

To quantify the suitability of an individual parameter combination for LSD RV extraction, we can compute the RMS of the resulting RV time series. We note, however, that the RMS may not be an optimal measure for RV purposes. For example, a planetary signal increases the RMS. Consequently minimising the RMS can lead to the undesirable suppression of the planetary signal in very flexible models. Also, the RMS can be inflated due to a trend in the RV time series. Such a trend or an RV offset between two observing seasons can arise due to e.g. a magnetic cycle. The effect of the latter on the absorption lines is wavelength-dependent \citep[e.g.][]{Reiners_2010} and thus different masks can lead to different long-term trends increasing the RMS without deteriorating the detection probability of short-period planets. To have a fair and meaningful comparison measure, we divide the RV time series in separate clusters using the KMeans algorithm as implemented in the Python library scikit-learn and compute the weighted average of the RMS of each cluster (cRMS). The weight of an individual cluster is set to the number of spectra therein.
For the target stars used in this study, the RMS and the cRMS are very well correlated with Pearson correlation coefficient greater than 0.87 for all stars. We therefore use the RMS for optimisation purposes of the LSD procedure. Whenever the final or components of the final LSD RV time series are displayed, we use the cRMS such that the results can be compared with the cRMS of the DRS-CCF method.

\end{itemize}

\subsection{Single-mask approach}  \label{sss:singlemaskapproach}
\begin{figure}
	\includegraphics[width=\columnwidth]{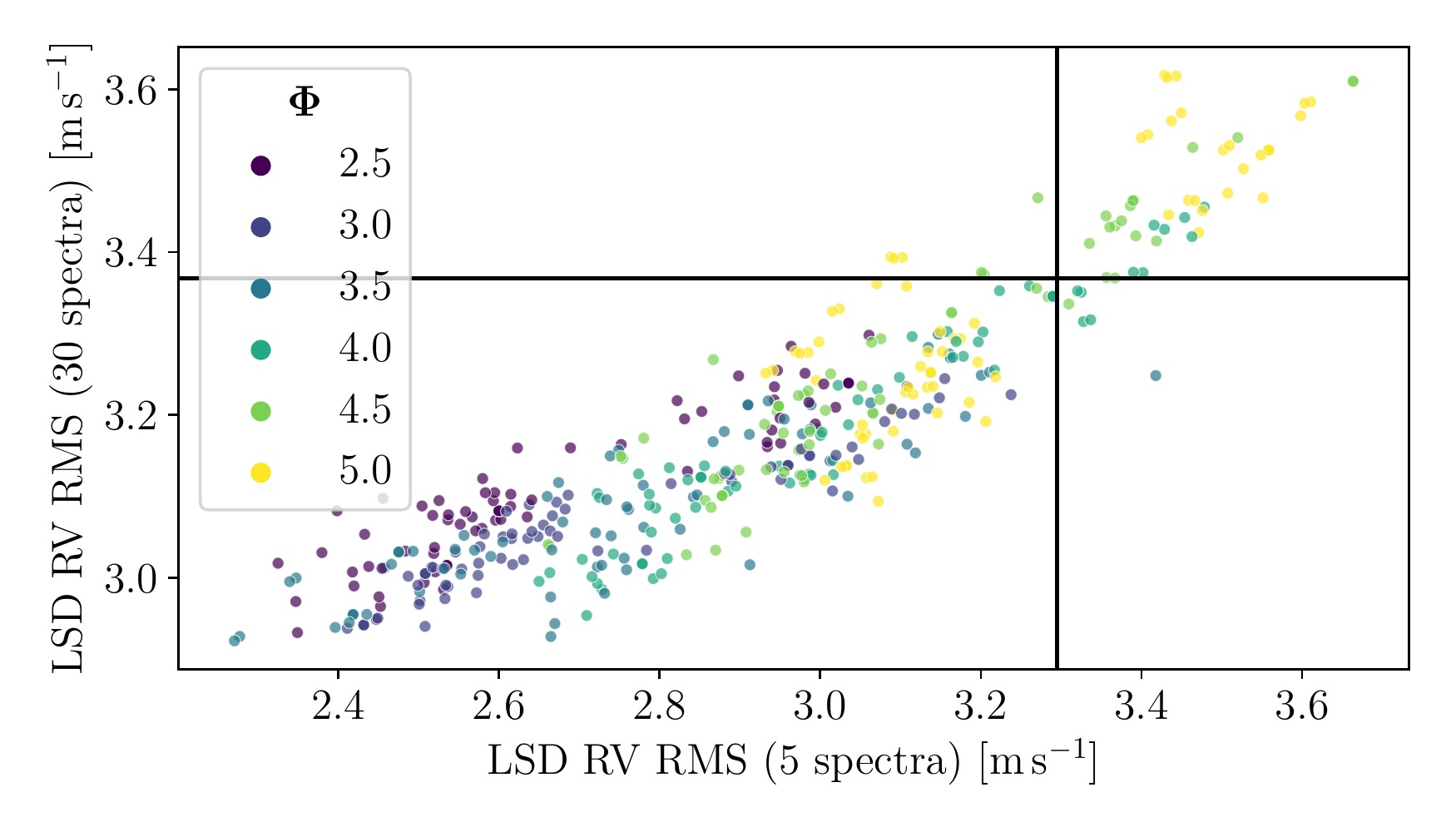}
    \caption{RV RMS of time series involving 5 spectra vs. time series consisting of 30 spectra of HD 4628 for different parameter combinations ($\mathbf{\Gamma}$, $\mathbf{\Xi}$, $\mathbf{\Phi}$, $\mathbf{\Theta}$). The RMS of the corresponding DRS CCF RV time series is indicated by black solid lines. The dependence on the velocity window size $\mathbf{\Phi}$ (in units of FWHM) is shown in this figure. Such an RMS--RMS correspondence cannot be established for all stars and thus we pursue a different parameter optimisation strategy (cf. Section \ref{ss:findparameters}).}
    \label{fig:rmsrmsphi}
\end{figure}

Given that about half a second is required to deconvolve a single spectrum, an extensive hyperparameter grid search is computationally very expensive. We thus tested whether suitable hyperparameters can be established based on a grid search on a subset of spectra. Ideally, this looks like Fig. \ref{fig:rmsrmsphi}, where we depict the RMS of the RV time series consisting of 5 spectra and 30 spectra for different parameter combinations. For the star at hand, a given parameter combination ($\mathbf{\Gamma}$, $\mathbf{\Xi}$, $\mathbf{\Phi}$, $\mathbf{\Theta}$) is favourable (unfavourable) for 30 spectra if it is also favourable (unfavourable) for 5 spectra. We could therefore conclude from only 5 spectra that a narrower velocity window $\mathbf{\Phi}$ is preferable for all spectra of this star.
It turns out, however, that finding a representative subset exhibiting such a correlation is not straightforward. We tested several different strategies such as selecting spectra that are close in time, equally spread out over time, very low or high in SNR, or equally distributed in SNR. None of these strategies consistently resulted in a clear RMS-RMS correlation, such as in Fig. \ref{fig:rmsrmsphi}, for all tested time series. There is tentative evidence that if there is a RMS-RMS correlation for about 10 hyperparameter combinations, the correlation persists for all hyperparameter combinations. However, in some cases, this correlation might be spurious and two subsets of spectra can have different optimal hyperparameters. In combination with the fact that the RMS is not an ideal quality measure, the single-best-combination approach can lead to various problems. 

\subsection{Multi-mask approach}  \label{sss:multimaskapproach}

For the reasons listed above, we opted to not rely on a single parameter combination but instead on several combinations simultaneously. We run the deconvolution code on all spectra for a coarse grid of 32 parameter combinations, which gives us 32 RV time series. We set the values of the individual parameters to any combination of $\mathbf{\Gamma}$ $\in$ \{0.5, 1.0\}, $\mathbf{\Xi}$ $\in$ \{0.8, 1.0\}, $\mathbf{\Phi}$ $\in$ \{2.5, 3.0, 3.5, 4.0\}, $\mathbf{\Theta}$ $\in$ \{0.1, 0.2\}.
The rationale for setting the parameters to these values is presented below.
\subsubsection{Hyperparameter grid}  \label{sss:paramgrid}
The cut-off for the model-spectrum deviation $\mathbf{\Gamma}$ was set to the values in \{0.5, 1.0\}. As apparent in Fig. \ref{fig:ms_deviation}, only about 0.05 per cent of the fluxes deviate by more than 1.0 from the convolution model, whereas 1 per cent deviate by more than 0.5. This means that fluxes that strongly deviate (i.e. by more than 1.0) are always excluded, while we explore the impact of a slightly more rigorous cut-off with $\mathbf{\Gamma}$ set to 0.5.

For the reasons outlined in Section \ref{sss:ap_lineselection}, we also explore how including or excluding the strongest lines in the VALD list impacts the RVs. First, we include all deep lines by setting $\mathbf{\Xi}$ to 1.0, then we run the RV extraction code again with $\mathbf{\Xi}$ set to 0.8 to exclude absorption lines with relative depth greater than 0.8.

As explained in Section \ref{ss:velocity_grid}, setting $\mathbf{\Phi}$ to 2.0 means that the velocity grid will be set to ($\text{RV}_0$-2.0$\cdot$HWHM, $\text{RV}_0$-HWHM+0.82, ... , $\text{RV}_0$+2.0$\cdot$HWHM), where $\text{RV}_0$ is the RV of the first spectrum in units of \kms and HWHM is the half width at half maximum of the first spectrum. Setting the parameter $\mathbf{\Phi}$ is not straightforward since it indirectly governs a few aspects. For instance, the grid width determines the fluxes that are excluded close to a pixel with high model-spectrum deviation (cf. Section \ref{sss:dd_lineselection}). 
If we had complete and accurate line information and absorption line blends were perfectly modelled by an additive model, the velocity grid width could be neglected. Since this is not the case and the LSD convolution model is essentially a superposition of scaled and shifted LSD common profiles, the least-squares minimisation can offset insufficiencies in the model by spuriously distorting the LSD common profile. The wider the velocity grid, the stronger neighbouring lines interfere with the convolution model of an individual line. Consequently, we opted for a more fine-grained scan of $\mathbf{\Phi}$ by testing the values in \{2.5, 3.0, 3.5, 4.0\}.

\begin{figure}
	\includegraphics[width=\columnwidth]{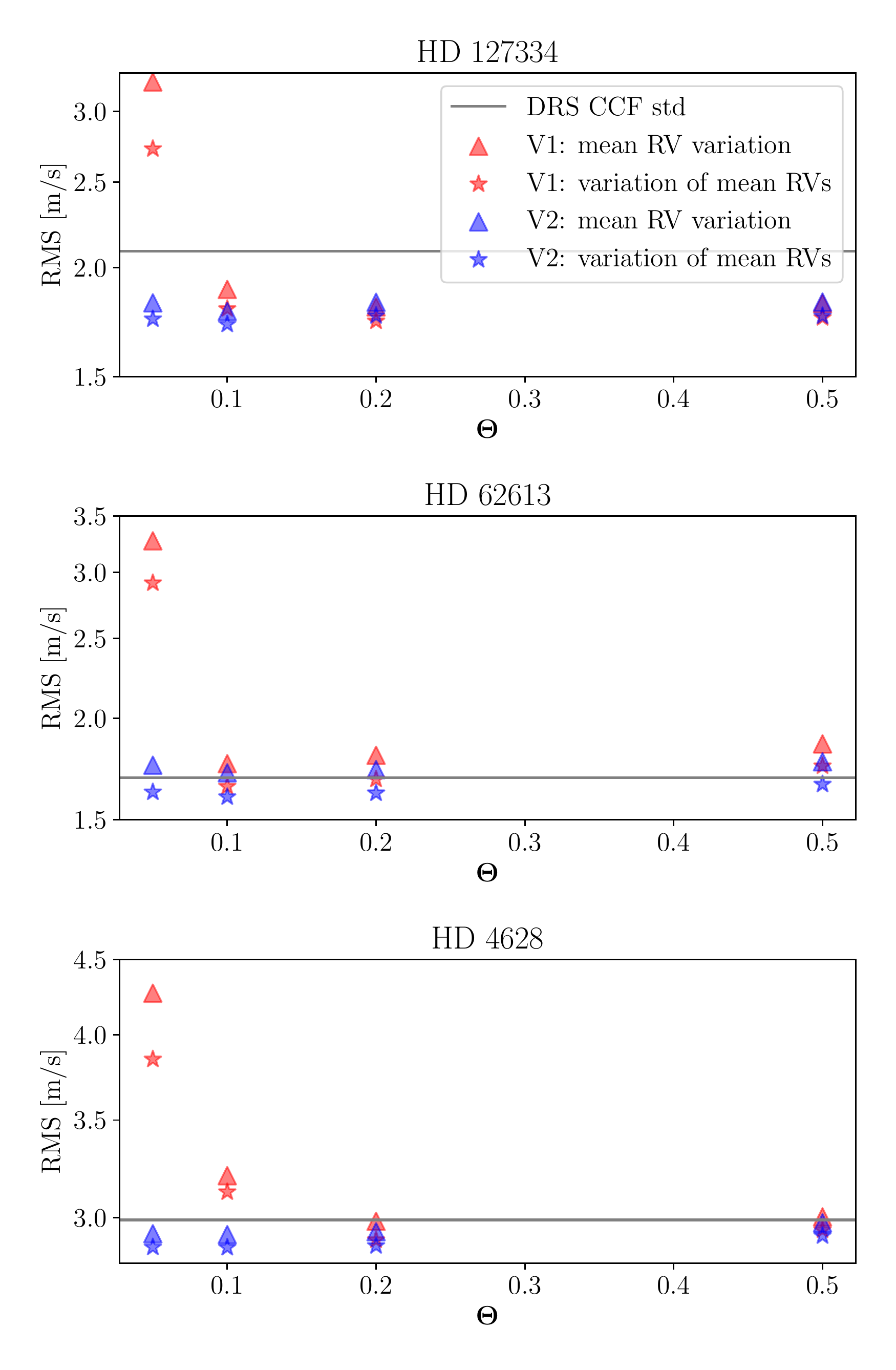}
    \caption{RMS of different RV time series depending on the telluric line depth threshold $\mathbf{\Theta}$. The mean of the standard deviations of the individual 16 RV time series are indicated with triangles resulting from running the code with any combination of $\mathbf{\Gamma}$ $\in$ \{0.5, 1.0\}, $\mathbf{\Xi}$ $\in$ \{0.8, 1.0\}, $\mathbf{\Phi}$ $\in$ \{2.5, 3.0, 3.5, 4.0\}. The stars show the standard deviation of the RV time series resulting from mean-combining the 16 individual time series using envelope weights.}
    \label{fig:vs1vsv2_combined}
\end{figure}

Lastly, we chose to restrict the telluric depth cut $\mathbf{\Theta}$ to \{0.1, 0.2\}. This choice was determined by testing the dependence of the RV time series scatter on $\mathbf{\Theta}$. To illustrate the importance of consistently including the same stellar absorption lines, we ran this test with both the telluric removal versions \textit{v1} and \textit{v2}.  As outlined in Section \ref{ss:tellurics}, \textit{v1} removes the same wavelengths in the observer's rest frame, whereas \textit{v2} consistently flags the same wavelengths in the barycentric reference frame.
We chose the three stars without any currently known planets (HD 127334, HD 62613, HD 4628) for this analysis and computed their RV time series for the 16 combinations of $\mathbf{\Gamma}$ $\in$ \{0.5, 1.0\}, $\mathbf{\Xi}$ $\in$ \{0.8, 1.0\}, $\mathbf{\Phi}$ $\in$ \{2.5, 3.0, 3.5, 4.0\}. Running this test for a number of parameter combinations is crucial to minimise the dependence of the RMS on the specific choice of the other hyperparameters. For each star, we computed the standard deviation of the 16 RV time series as well as the standard deviation of the time series resulting from mean-combining these 16 time series. In Fig. \ref{fig:vs1vsv2_combined}, we show the mean of the former (triangles) and compare with the latter (stars) for both \textit{v1} and \textit{v2}.  Firstly, we note that mean-combining the 16 individual RV time series always leads to a new RV time series with a lower RMS compared to the average time series. Secondly, we can see from Fig. \ref{fig:vs1vsv2_combined}, that setting a restrictive value for $\mathbf{\Theta}$ leads to very high scatter for \textit{v1}, while \textit{v2} still performs well. This corroborates the explanation for the necessity of \textit{v2} given in Section \ref{ss:tellurics}. 
We note, however, that setting $\mathbf{\Theta}$ to a very low value leads to the exclusion of valuable data and can increase the scatter for \textit{v2} as well, as visible in Fig. \ref{fig:vs1vsv2_combined} for $\mathbf{\Theta}$ set to 0.05.

Given that $\mathbf{\Theta}$ = 0.1 leads to the lowest scatter for all three stars, we include $\mathbf{\Theta}$ = 0.1 in the parameter grid. To further explore the impact of a less restrictive telluric line removal approach, we also include $\mathbf{\Theta}$ = 0.2.

\subsubsection{Final RVs}  \label{sss:combine_rvs}
We subsequently compute the final LSD RV time series by mean-combining the time series that have an RMS below the median RMS of all time series (i.e. 16 out of 32 RV time series). This ensures that unsuitable parameter combinations leading to the inclusion of variable lines are excluded, while not relying on a single parameter combination. The reasons for averaging 16 out of 32 RV time series is further explained in the following.

Fluxes that deviate significantly from the convolution model can distort the common profile because we minimise the deviation between the model and the spectrum to find the common profile. Since the convolution model cannot reproduce these regions of the spectrum, it can only track the time-evolution of these regions to a limited degree. Consequently, these regions can induce RV scatter. 

Data points with extreme model-spectrum deviation greater than 1 are excluded for all hyperparameter combinations. This leads to the removal of a marginal number of data points. However, about 1 per cent of all fluxes deviate between 0.5 and 1 from the convolution model, as shown in Fig. \ref{fig:ms_deviation}. By removing these data points, we risk discarding valuable RV information. The multi-mask approach provides us with the necessary flexibility to tackle this issue.
 
Only half of all hyperparameter combinations include these problematic regions of the spectrum (any combination with $\mathbf{\Gamma}$ = 1.0). Since we select the 16 time series with the lowest scatter, all the time series computed with $\mathbf{\Gamma}$ = 1.0 get discarded if they lead to very high scatter independent of the other hyperparameters. Conversely, if excluding these regions removes RV information and thus increases the scatter, the other 16 time series are discarded.

We do not expect either of these cases to be the standard outcome. Instead, we expect that the extent of the problem depends equally on the depth of the included lines $\mathbf{\Xi}$ or, to a lesser degree, on the width of the velocity grid $\mathbf{\Phi}$.

Spectral regions with deep lines are likely to get an undeservedly high weight since we expect a deep spectral line. These flux values are only included in 8 out of the 32 RV time series (any combination with $\mathbf{\Gamma}$ = 1.0 and $\mathbf{\Xi}$ = 1). By computing the RV time series on a parameter grid, we thus probe the impact of a more or less strict quality cut-off using various parameters and the combinations thereof.

As mentioned in Section \ref{sss:singlemaskapproach}, there is often no single hyperparameter combination that is optimal for all subsets of spectra. 
By mean-combining 16 time series, we mitigate this problem since if a specific parameter combinations leads to increased scatter for a subset of spectra, the RVs from the other included time series act as a counterweight. As a further advantage, the 16/32-mean is significantly less prone to being artificially distorted to reduce the RMS as compared to the time series with the lowest RMS.

\begin{figure}
	\includegraphics[width=\columnwidth]{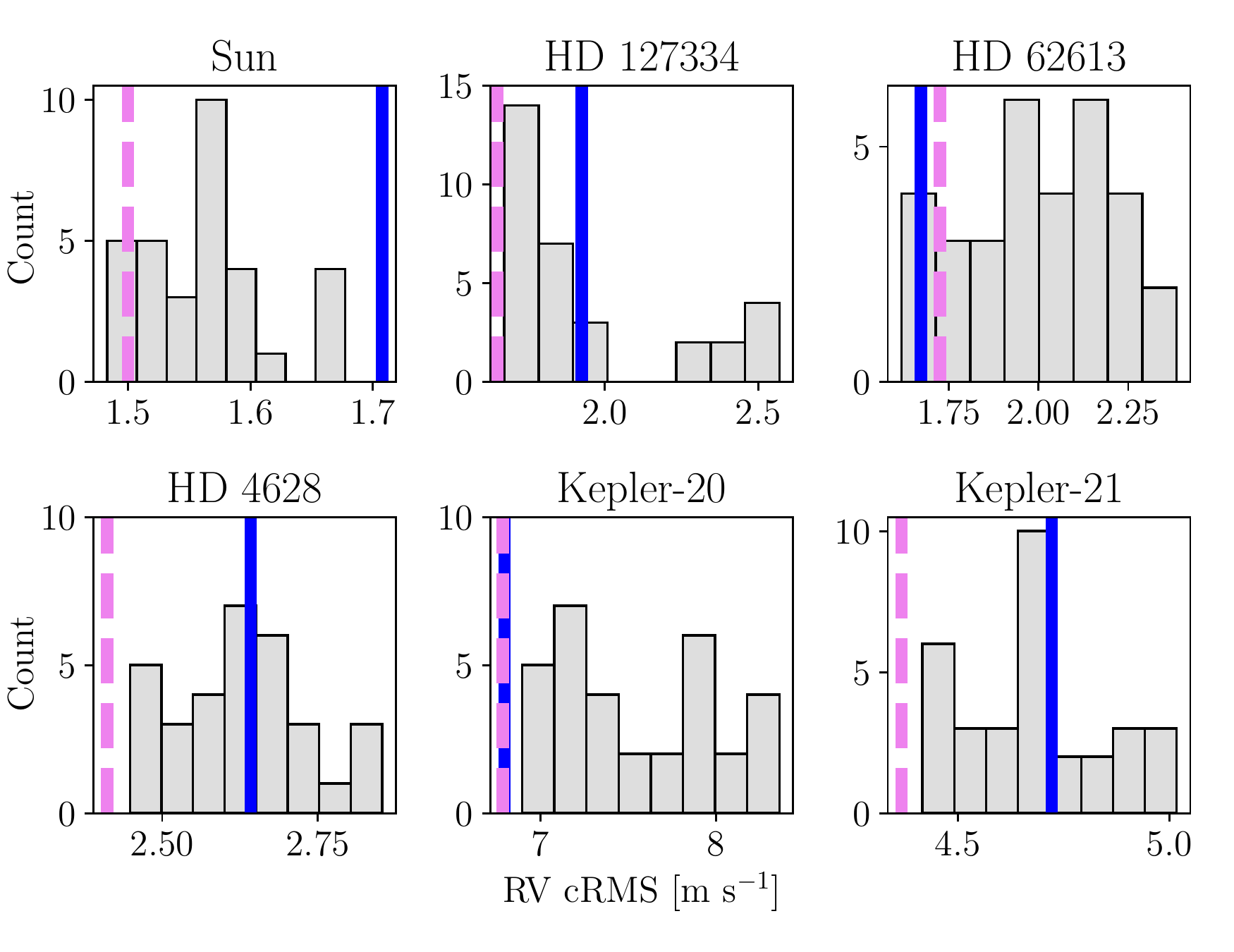}
    \caption{Histogram of the cRMS of the 32 time series produced using the 32 different parameter combinations using flux weights. The blue vertical line indicates the cRMS of the DRS CCF RVs while the pink line shows the cRMS of the RV time series computed by mean-combining the 16 RV time series with RMS below the median.}
    \label{fig:rms_paramdep_all}
\end{figure}

Fig. \ref{fig:rms_paramdep_all} shows how the cRMS for all 32 parameter combinations (grey histogram) compares with the cRMS of the DRS CCF method and the mean-combination of the 16 RV time series (16/32-mean) with lowest scatter. In line with the above arguments, we find that the 16/32-mean time series is significantly less marked by scatter than the average time series and always comparable or better than the best single parameter combination.

To characterise the RV variation due to the hyperparameter choice, we analysed the distribution of the RVs resulting from the parameter combinations for each spectrum. From the Shapiro-Wilk test \citep{Shapiro_1965}, we conclude that these RVs are normally distributed for about 85 per cent of the spectra with a typical standard deviation per spectrum of about 0.5 m/s for HD 127334, HD 62613, and HD 4628. The standard deviation of these RVs (for the same spectrum) can thus serve as an additional RV uncertainty component and shows whether the extracted RV is highly dependent on the parameter combination.

\begin{figure}
	\includegraphics[width=\columnwidth]{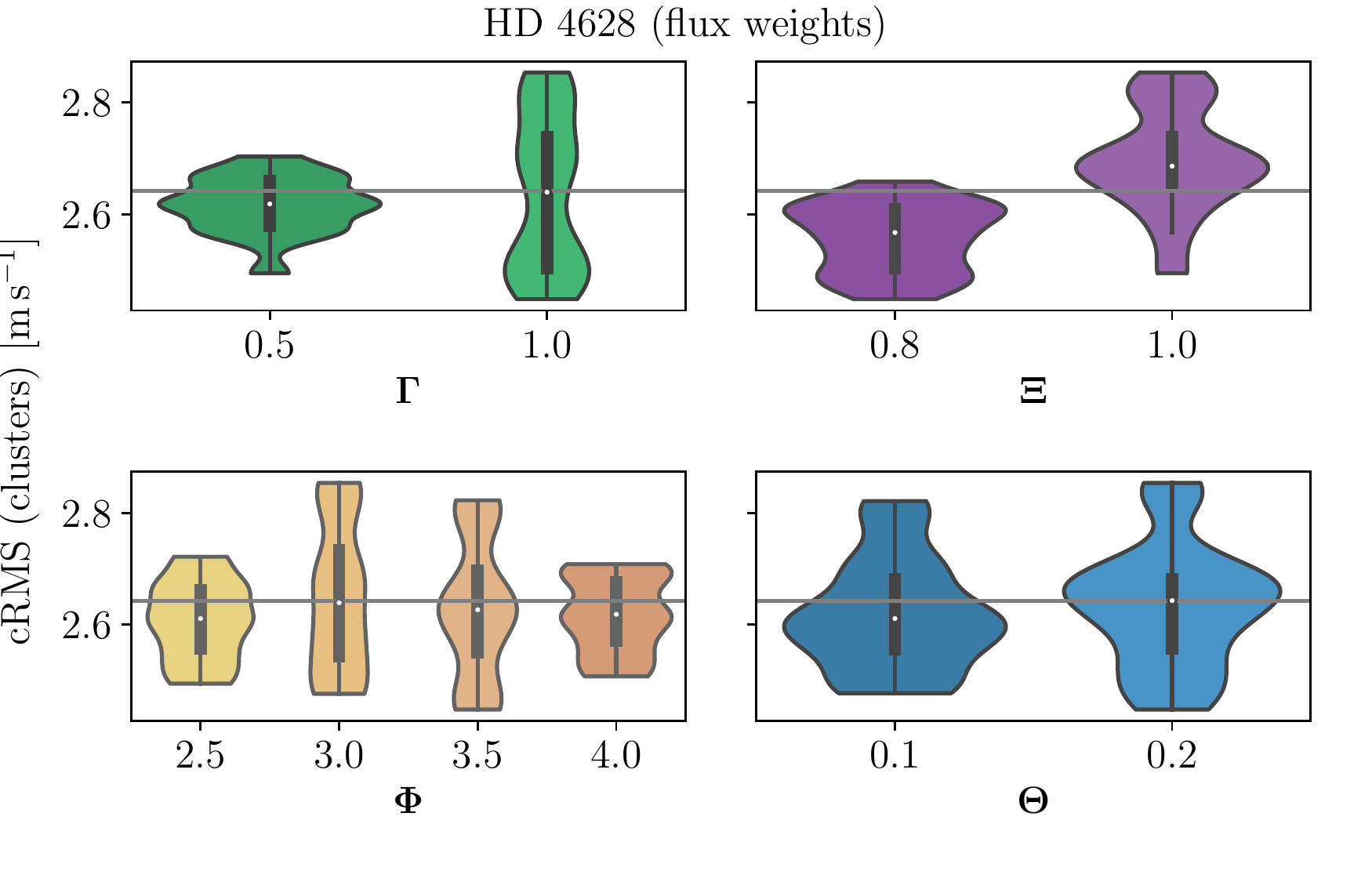}
	\includegraphics[width=\columnwidth]{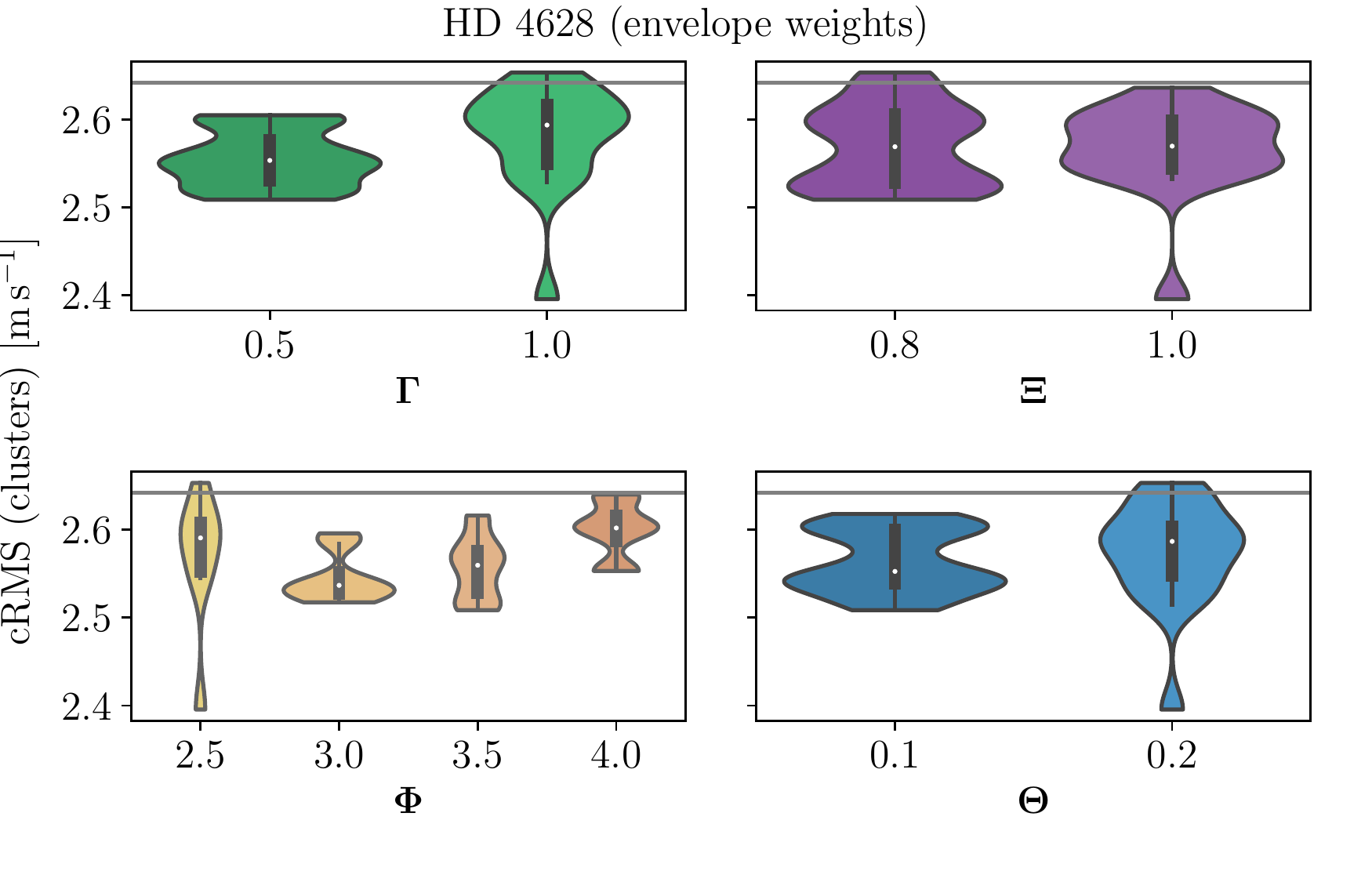}
    \caption{Violin plot showing the RMS distribution (weighted mean cluster RMS, cRMS) of the 32 LSD RV time series split by the varied parameters (cf. definitions in Section \ref{ss:findparameters}) for HD 4628. The horizontal line shows the equivalent scatter measure for the DRS CCF time series. The top four panels show the RMS distribution for flux weights, while the bottom four panels show the same for envelope weights.}
    \label{fig:rms_paramdep}
\end{figure}

Fig. \ref{fig:rms_paramdep} shows how the RMS of the LSD RV time series depends on the individual parameters. The top left subplot of both panels shows that excluding areas with high model-spectrum deviation is advantageous reducing the scatter in the RV time series.
Setting the relative depth threshold to 0.8 (top right subplot) and thus excluding the deepest lines is beneficial if we use flux weights but it hardly has any positive impact if we use envelope weights. The same behaviour is present for other stars as visible in the figures displayed in Appendix \ref{a:rmsdeponparameters}. This indicates that in the context of a time series, using flux weights is putting too much weight on deep lines (cf. Section \ref{ss:weight_matrix}).

\section{Results} \label{s:results}

We tested the LSD pipeline on both the Sun and a diverse set of FGK-type stars to assess its performance. As a measure of the performance of our LSD extraction pipeline, we compute the RMS, MAD and cRMS (as defined in Section \ref{ss:findparameters}) of the full RV time series. Furthermore, we compute the periodograms for a star with a single known planet and a known multi-planet system (cf. Section \ref{sss:keplertest}). The results can be found in Table \ref{impr_per_step_stars}. For each star, we list the scatter measures for the DRS CCF time series in line (a). The results in line (b) refer to the RVs resulting from only one parameter combination ($\mathbf{\Gamma}$ = $\infty$, $\mathbf{\Xi}$ = 1, $\mathbf{\Phi}$ = 3, $\mathbf{\Theta}$ = $\infty$) with flux weights and no telluric line correction. Most of this scatter is removed by dividing by the static telluric model and by masking telluric lines deeper than 0.5 (c). Further improvements through the use of multiple tailored masks and \textit{telluric map v2} lead to the results in line (d), (e), and (f). The approximate improvement due to the LSD method itself with some minor enhancements such as using RASSINE (Section \ref{s:Continuum_correction}) and suitable order weights (Section \ref{ss:order_weighting}) can be estimated by comparing lines (c) and (a). The improvement due to the multi-mask approach can be deduced by comparing lines (c) and (d).

We regard the RV results using flux weights (printed in bold) as our main results and all additional analysis is based on these time series. Note that the MAD of the final LSD time series is lower for the LSD RVs (d) as compared to the CCF RVs (a) for all stars, while the RMS and the cRMS are lower or equal for all stars except HD 62613 and significantly lower for the Sun, HD 127334, HD 4628, and Kepler-21.
We further investigate the individual RV time series in the following subsections.

\subsection{Sun}

The LSD pipeline was tested on 300 solar spectra from the set of observations presented in \citet{Dumusque_2021}. The three scatter measures described above are computed in the heliocentric frame of reference (cf. Table \ref{impr_per_step_stars}). The heliocentric correction was assumed to be unknown throughout the computation of the RV time series, however. The RMS of the time series in the heliocentric frame only captures the scatter due to stellar activity and e.g. photon noise without the RV effect of any planets and is thus a unique tool for the analysis of an RV extraction code. As apparent in Table \ref{impr_per_step_stars}, the LSD pipeline produces RV time series that show less scatter compared to the DRS CCF technique for all three weighting schemes. The model assumptions are well met by the high-quality solar spectra, hence flux weights are clearly favoured.

\subsection{HARPS-N standard stars without known planets}

The solar data set consists of very high signal-to-noise spectra at high cadence of a star that is resolved in the sky, which sets it apart from nightly HARPS-N observations of other stars. We thus ran the code on some standard HARPS-N targets without any known planets and computed the same quality measures as for the Sun.
The results in Table \ref{impr_per_step_stars} show that the different weighting schemes (cf. Fig. \ref{fig:upperenvelope}), do not improve the scatter measures homogeneously for all stars.
The RVs extracted using envelope weights have a lower cRMS than the DRS CCF RVs for all three stars. Using uniform weights, on the other hand, reduces the cRMS for two out of the three targets. The RV time series of HD 62613 computed using flux weights has a slightly higher RMS of 1.75\ms as compared to the RMS of 1.69\ms of the DRS-CCF RV time series. This is due to one spectrum with very low SNR of 32 at 550 nm (median SNR: 147) leading to a slight outlier. Without this low-quality spectrum, the RMS of the LSD RVs is 1.62\ms, while the RMS of the DRS CCF RVs is 1.68\ms. The same spectrum does not lead to a deviating RV if we use envelope errors or uniform errors which indicates that these weightings can outperform the standard flux weights when the model assumptions, such as the underlying noise being Gaussian, are not met. Since the airmass of the HD 62613 observations is high (median: 1.64), we expect residual scatter due to tellurics and thus a more sophisticated telluric transmittance model is expected to reduce the RV RMS.

\begin{table}
\caption{RMS, MAD, and cRMS (the weighted mean of the RMS of observation clusters close together in time) of the 6 HARPS-N targets. Line (a) shows the scatter measures for the DRS-CCF time series. 
Line (b) refers to the RVs resulting from the parameter combination ($\mathbf{\Gamma}$ = $\infty$, $\mathbf{\Xi}$ = 1, $\mathbf{\Phi}$ = 3, $\mathbf{\Theta}$ = $\infty$) using flux weights and no telluric line correction.
In line (c), we list the scatter of the time series if we apply LSD with only one parameter combination, without excluding lines with high model-spectrum deviation, masking tellurics deeper than 0.5 using \textit{telluric map v1}. In lines (d), (e), and (f), we list the scatter measures of the RV time series computed from the better half of the parameter combinations for different weighting schemes.}
\label{impr_per_step_stars}

\begin{tabular}{lrrr}
\hline
\hline
Sun  & RMS & MAD & cRMS\\
    & \msnw & \msnw & \msnw \\
\textbf{(a) DRS-CCF} & 2.07 & 1.25 & 1.71\\
(b) LSD All lines      &8.71 &7.23 &8.53 \\
(c) LSD All lines (\textit{v1}, $\mathbf{\Theta}$ = 0.5)     &1.68 &1.09 & 1.60\\
(d) \textbf{LSD 16/32-mean flux weights} &1.67 & 1.12 & 1.50 \\
(e) LSD 16/32-mean envelope weights    &1.90 & 1.29 & 1.61\\
(f) LSD 16/32-mean uniform weights & 1.78 & 1.16 & 1.61\\
\hline
\hline
\end{tabular}

\begin{tabular}{lrrr}
HD 127334  & RMS & MAD & cRMS\\
    & \msnw & \msnw & \msnw \\
\textbf{(a) DRS-CCF} & 2.09 & 0.92 & 1.93\\
(b) LSD All lines      &23.80 &14.16 &19.72 \\
(c) LSD All lines (\textit{v1}, $\mathbf{\Theta}$ = 0.5)     &2.23 &1.18 &2.15 \\
(d) \textbf{LSD 16/32-mean flux weights}    &1.76 & 0.90 & 1.65  \\
(e) LSD 16/32-mean envelope weights    &1.71 & 0.88 & 1.61\\
(f) LSD 16/32-mean uniform weights & 1.71 & 0.81 & 1.62\\
\hline
\hline
\end{tabular}

\begin{tabular}{lrrr}
HD 62613  & RMS & MAD & cRMS\\
    & \msnw & \msnw & \msnw \\
\textbf{(a) DRS-CCF} & 1.69 & 1.15& 1.67\\
(b) LSD All lines    &26.87 &22.20 &24.92 \\
(c) LSD All lines (\textit{v1}, $\mathbf{\Theta}$ = 0.5)     &2.19 &1.31 & 2.17\\
(d) \textbf{LSD 16/32-mean flux weights}    & 1.75 & 1.08 & 1.73  \\
(e) LSD 16/32-mean envelope weights    & 1.56 & 1.05 & 1.55\\
(f) LSD 16/32-mean uniform weights &  1.84 & 1.25 & 1.82\\
\hline
\hline
\end{tabular}

\begin{tabular}{lrrr}
HD 4628  & RMS & MAD & cRMS \\
    & \msnw & \msnw & \msnw \\
\textbf{(a) DRS-CCF} & 2.99&2.44&2.64\\
(b) LSD All lines    & 19.33 &9.16 &18.97 \\
(c) LSD All lines (\textit{v1}, $\mathbf{\Theta}$ = 0.5)     &3.21 &1.86 & 2.97\\
(d) \textbf{LSD 16/32-mean flux weights}    &2.69 & 1.96 & 2.41 \\
(e) LSD 16/32-mean envelope weights    &2.80 & 1.87 & 2.47 \\
(f) LSD 16/32-mean uniform weights & 2.54 & 1.82 & 2.21\\
\hline
\hline
\end{tabular}

\begin{tabular}{lrrr}
Kepler-20  & RMS & MAD & cRMS\\
    & \msnw & \msnw & \msnw \\
\textbf{(a) DRS-CCF} & 6.94 & 5.02 & 6.79\\
(b) LSD All lines      &12.59 & 7.90& 13.07 \\
(c) LSD All lines (\textit{v1}, $\mathbf{\Theta}$ = 0.5)     &8.31 &5.84 & 8.08\\
(d) \textbf{LSD 16/32-mean flux weights}    & 6.87 & 4.75 & 6.79 \\
(e) LSD 16/32-mean envelope weights    & 7.07 & 4.92 & 6.99\\
(f) LSD 16/32-mean uniform weights & 7.23 & 5.38 & 7.18\\
\hline
\hline
\end{tabular}

\begin{tabular}{lrrr}
Kepler-21  & RMS & MAD & cRMS\\
    & \msnw & \msnw & \msnw \\
\textbf{(a) DRS-CCF} & 6.10 & 4.05 & 4.72\\
(b) LSD All lines      &6.35 &4.38 &5.96 \\
(c) LSD All lines (\textit{v1}, $\mathbf{\Theta}$ = 0.5)     &5.19 &3.70 & 4.50\\
(d) \textbf{LSD 16/32-mean flux weights} & 5.03 & 3.48 & 4.37\\
(e) LSD 16/32-mean envelope weights & 5.20 & 3.23 & 4.52\\
(f) LSD 16/32-mean uniform weights & 5.29 & 3.55 & 4.80\\
\hline
\hline
\end{tabular}

\end{table}

\subsection{Transiting planet hosts} \label{sss:keplertest}

Lastly, we tested the LSD pipeline on stars with published planets. We chose two Kepler targets, Kepler-20 and Kepler-21. For Kepler-20, two spectra were excluded from the analysis as they led to outlier RVs for both the DRS CCF and the LSD method.

\begin{figure*}
	\includegraphics[width=2.\columnwidth]{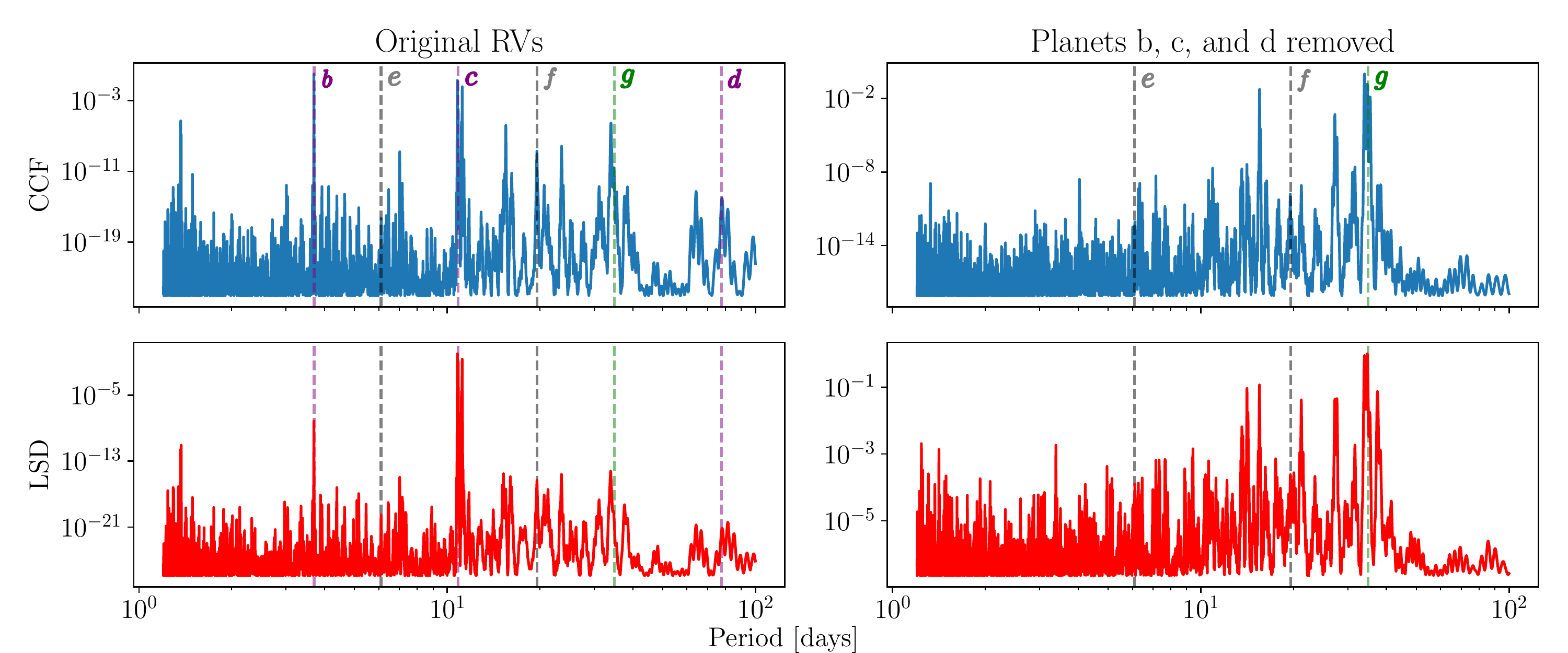}
    \caption{BGLS periodogram for Kepler-20 time series using the DRS CCF method (top) or the LSD pipeline (bottom). For the periodograms on the left, the RVs without any manipulation were analysed. On the right, we show the periodograms for the RVs with the signals associated with planets b, c, and d removed.
    The transiting planets \citep{Borucki_2011,Gautier_2012,Fressin_2012} are indicated by vertical dashed purple lines or grey lines if they have no mass measurement, while the period attributed to a non-transiting planet in \citet{Buchhave_2016} is marked by a green dashed line.}
    \label{fig:kepler20_periodogram}
\end{figure*}

Kepler-20 hosts a multi-planet system \citep{Borucki_2011,Gautier_2012,Fressin_2012}, which means that a periodogram is not an ideal analysis tool since many Keplerian signals are superimposed, but it can still serve to compare the dominant signals in the two time series. In Fig. \ref{fig:kepler20_periodogram}, we show the Bayesian Generalised Lomb-Scargle (BGLS) Periodogram \citep{Mortier_2015} for Kepler-20.
The BGLS periodogram was first computed for the original RVs for both the LSD and the CCF RVs (left panel). The transiting planets b (3.7 d), c (10.9 d), and d (77.6 d) are clearly visible for both techniques, while planets e (6.1 d) and f (19.6 d) show no prominent peaks. To improve the visibility of other signals, we then modelled the three transiting planets b, c, and d and subtracted them from the original RVs.
To generate this model, we used the \texttt{PyORBIT} software \citep{Malavolta_2016} with the \texttt{MultiNest} sampler \citep{Feroz_2008,Feroz_2009,Feroz_2019}, with 1000 live points. We assumed circular orbits for all three included planets, and constrained the periods and transit timings to the values listed in \citet{Buchhave_2016}. After subtracting these sinusoids from the original RVs, we generate another BGLS periodogram displayed on the right in Fig. \ref{fig:kepler20_periodogram}. Without the dominant sinusoidal signals of planets b, c, and d, we see another periodogram peak emerging (LSD RVs: 34.7 d, CCF RVs: 34.0 d), which agrees with the period of planet g (34.9 d) predicted in \citet{Buchhave_2016}. However, we note that this signal could still be generated or amplified by magnetic activity (see \citep{Nava_2020} for a review).

\begin{figure}
	\includegraphics[width=\columnwidth]{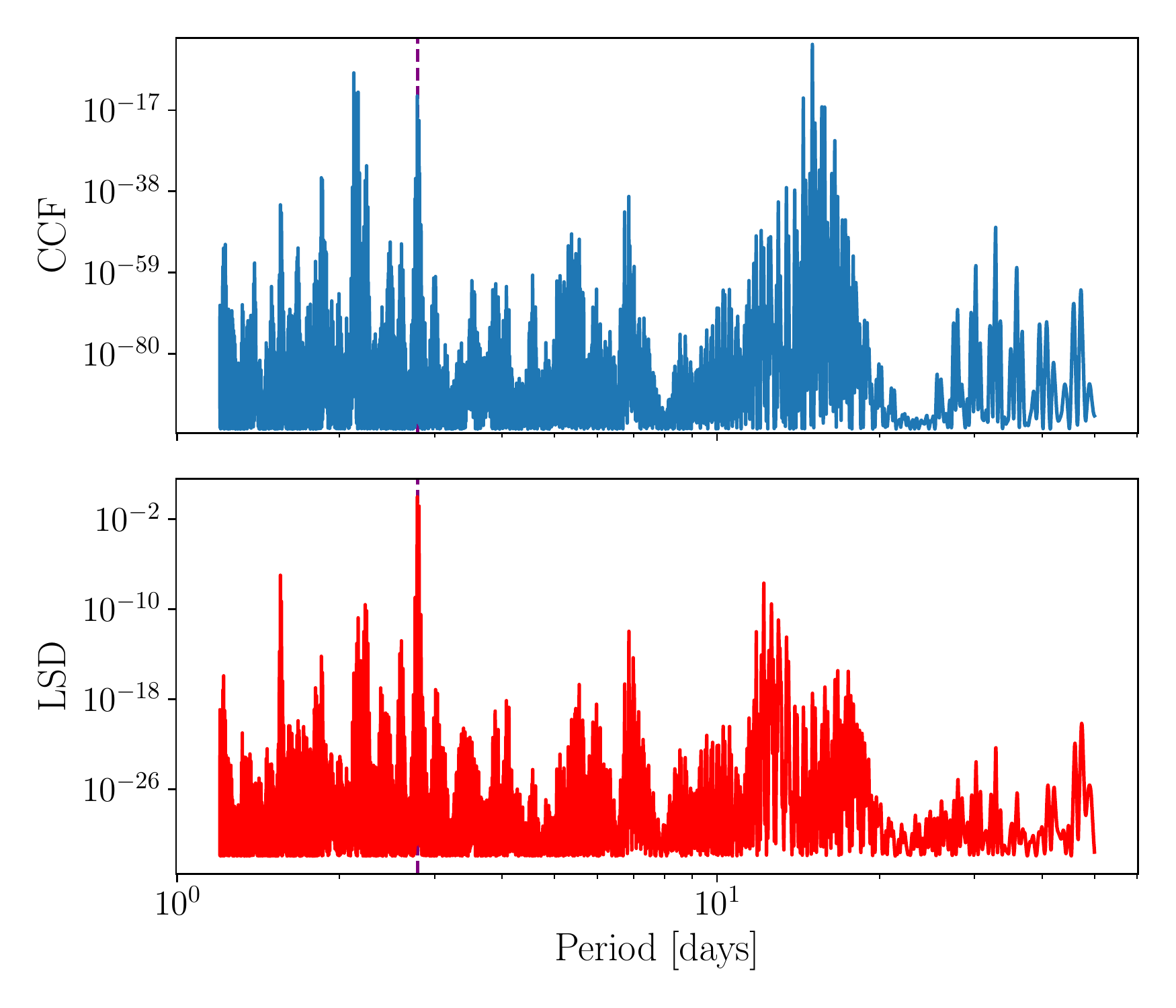}
    \caption{BGLS periodogram for Kepler-21 time series using the DRS CCF method (top) or the LSD pipeline (bottom). The signal of Kepler-21b at 2.8 days \citep{Howell_2012,Lopez_Morales_2016} indicated by the vertical dashed violet line is stronger in the LSD RV periodogram while the forest of peaks at longer periods is suppressed.}
    \label{fig:kepler21_periodogram}
\end{figure}

Fig. \ref{fig:kepler21_periodogram} shows the BGLS periodograms for Kepler-21 which has a single known transiting planet \citep{Howell_2012,Lopez_Morales_2016}. This planet's signal is clearer in the LSD periodogram, while the forest of longer period peaks around 15 days, attributed to stellar activity in \citet{Lopez_Morales_2016}, is clearly less prominent as compared to the CCF periodogram.

\section{Conclusion and Outlook} \label{s:conclusion_outlook}

We have built and tested an RV extraction pipeline applying LSD to high-resolution stellar intensity spectra of FGK-type stars using multiple tailored masks. This pipeline normalises S2D echelle order spectra, masks telluric absorption lines, and computes the RV over a range of parameters. During this process, special care is taken to avoid introducing spurious RV shifts due to intrinsic model insufficiencies or incomplete knowledge.

As shown in this study, the multitude of instrumental, atmospherical, and stellar effects on the RV measurement can be balanced out better by using multiple tailored masks instead of one single mask. We also show that it can be detrimental to assume that the flux uncertainty for an assumed stellar spectrum model is given by photon and instrumental errors of the recorded spectrum. The latter is due to the presence of imperfectly corrected telluric absorption lines, potentially unknown instrumental effects, and stellar activity varying the profile of an absorption line over time (cf. Section \ref{ss:weight_matrix}). These factors do not impact all stars equally, hence we recommend to test envelope weights as well as the standard flux weights for a given target.

The pipeline outputs RVs, RV uncertainties, and common profiles with uncertainties for each spectrum. The LSD pipeline has been shown to produce RV time series with generally lower scatter compared to the CCF method, while retaining the planetary RV signal at a comparable or better level.
In its current state, the pipeline masks and partly corrects telluric absorption lines using a static transmittance spectrum. This approach is robust and fast, but disregards the atmospheric conditions and the target's airmass at the time of the individual observations. Providing tailored telluric transmittance spectra to the pipeline is thus expected to further reduce the RV scatter.

\subsection{Application}

The common profiles produced by the LSD pipeline can be fed to codes such as {\sc SCALPELS} \citep{CollierCameron_2021} or machine learning techniques such as presented in \citet{deBeurs_2020} to further correct for stellar activity. Furthermore, the convolution model of the stellar spectrum allows to easily measure the variability of stellar absorption lines and identify stellar or instrumental effects using post-processing techniques such as those applied in {\sc YARARA} \citep{Cretignier_2021}. We thus see the presented code as a standalone RV extraction code as well as a base for other codes and extensions.

In addition to HARPS-N data, we have also tested the LSD pipeline on ESPRESSO and EXPRES data, yielding comparable results to the standard pipelines. For EXPRES data, minor adjustments to the code's data read-in are necessary. We only include the HARPS-N results for the in-depth performance analysis in this work.

\subsection{Future work}

The presented core LSD algorithm can be expanded to include further improvements in the RV extraction. As a next step, we will include a standard telluric line correction within the code. As it stands, deep telluric absorption lines are excluded and partly divided out. Using the {\sc Wobble} code \citep{Bedell_2019} is likely to yield a good data-driven telluric transmittance model. 

Furthermore, the variability of each line with respect to the LSD model can be used as an indicator for unsuitable lines, but also to detect instrumental effects.

Finally, in our upcoming work, the code will be extended to extract the magnetic flux from the individual lines via Zeeman broadening or Zeeman intensification \citep{Kochukhov_2020}. This is a very promising avenue to mitigating stellar activity as shown with solar observations \citep{Haywood_2016,Haywood_2020}. Having direct access to magnetic flux variation derived from intensity spectra can open the door to extract the small signals of Earth-like planets orbiting solar-type stars.

\section*{Acknowledgements}

FL gratefully acknowledges a scholarship from the Fondation Zd\u{e}nek et Michaela Bakala. FL would like to thank Belinda Nicholson for useful discussions. AM acknowledges support from the senior Kavli Institute Fellowships. ACC acknowledges support from STFC consolidated grant numbers ST/R000824/1 and ST/V000861/1.
 The HARPS-N project has been funded by the Prodex Program of the Swiss Space Office (SSO), the Harvard University Origins of Life Initiative (HUOLI), the Scottish Universities Physics Alliance (SUPA), the University of Geneva, the Smithsonian Astrophysical Observatory (SAO), the Italian National Astrophysical Institute (INAF), the University of St. Andrews, Queen’s University Belfast, and the University of Edinburgh.

\section*{Data Availability}

The solar spectra are available on the Data \& Analysis Center for Exoplanets (DACE) web-platform operated by the University of Geneva (\url{https://dace.unige.ch/sun/?}). 



\bibliographystyle{mnras}
\bibliography{MM-LSD} 




\appendix
\section{Analytic expression for common profile} \label{app:Zderivation}
To find the best-fitting LSD common profile (cf. Section \ref{ss:stokes_i_lsd}, Eq. \ref{eq:minimisation} and \ref{eq:deconvolution}), we minimise the expression below. The matrix $\textbf{M}$ is generated from the wavelength and depth information for the absorption lines from VALD3. The number of rows in $\textbf{M}$ is equal to the number of flux measurements, while the number of columns is equal the number of velocity grid points of the LSD common profile $\text{Z}$. The quadratic matrix \textbf{S} contains the inverse squared uncertainties of the flux measurements on the diagonal and zeros off the diagonal. First, we expand the minimisation expression.

\begin{align}
\chi ^2
&=  (\text{Y} - \textbf{M} \text{Z})^\intercal \textbf{S} (\text{Y} - \textbf{M} \text{Z})\\
&= (\text{Y}^\intercal - \text{Z}^\intercal \textbf{M}^\intercal) \textbf{S} (\text{Y} - \textbf{M}\text{Z})\\
&= \text{Y}^\intercal \textbf{S} \text{Y} - \text{Y}^\intercal \textbf{S} \textbf{M} \text{Z}- \text{Z}^\intercal \textbf{M}^\intercal \textbf{S} \text{Y}+\text{Z}^\intercal \textbf{M}^\intercal \textbf{S} \textbf{M} \text{Z}\\
&=  \text{Y}^\intercal \textbf{S} \text{Y} - 2\text{Y}^\intercal \textbf{S} \textbf{M} \text{Z}+\text{Z}^\intercal \textbf{M}^\intercal \textbf{S} \textbf{M} \text{Z}
\end{align}
Where we used that $\text{Z}^\intercal \textbf{M}^\intercal \textbf{S} \text{Y}$ is symmetric and thus $\text{Z}^\intercal \textbf{M}^\intercal \textbf{S} \text{Y}=(\text{Z}^\intercal \textbf{M}^\intercal \textbf{S} \text{Y})^\intercal = \text{Y}^\intercal \textbf{S} \textbf{M} \text{Z}$ in the last step (\textbf{S} is diagonal and thus $\textbf{S}^\intercal = \textbf{S}$).\\
Therefore, we must minimise the above term:
\begin{align}
0 &= \nabla_{\text{Z}} (\text{Y}^\intercal \textbf{S} \text{Y} - 2\text{Y}^\intercal \textbf{S} \textbf{M} \text{Z}+\text{Z}^\intercal \textbf{M}^\intercal \textbf{S} \textbf{M} \text{Z}) \label{ztmtmz}\\
&= - 2\text{Y}^\intercal \textbf{S} \textbf{M} +\text{Z}^\intercal \textbf{M}^\intercal \textbf{S} \textbf{M} +\text{Z}^\intercal \textbf{M}^\intercal \textbf{S} \textbf{M} 
\end{align}
In the last step we used that $\text{Z}^\intercal \textbf{M}^\intercal \textbf{S} \textbf{M} \text{Z}$ is a quadratic form. If $\alpha$ is a quadratic form
\begin{equation}
    \alpha = \textbf{x}^\intercal\textbf{A}\textbf{x},
\end{equation}
it holds that
\begin{equation}
    \frac{\partial\alpha}{\partial\textbf{x}} = \textbf{x}^\intercal(\textbf{A}+\textbf{A}^\intercal).
\end{equation}
Consequently, we must solve the following equation:
\begin{equation}
    \text{Y}^\intercal \textbf{S} \textbf{M} = \text{Z}^\intercal \textbf{M}^\intercal \textbf{S} \textbf{M}
\end{equation}
or
\begin{equation} 
    \textbf{M}^\intercal \textbf{S} \text{Y} = \textbf{M}^\intercal \textbf{S} \textbf{M}\text{Z}. \label{eq:vtm}
\end{equation}
We can solve Eq. \ref{eq:vtm} directly via $\text{Z} =   (\textbf{M}^\intercal \textbf{S} \textbf{M})^{-1}\textbf{M}^\intercal \textbf{S} \text{Y}$.

\section{RMS dependence on parameters} \label{a:rmsdeponparameters}

\begin{figure}
	\includegraphics[width=\columnwidth]{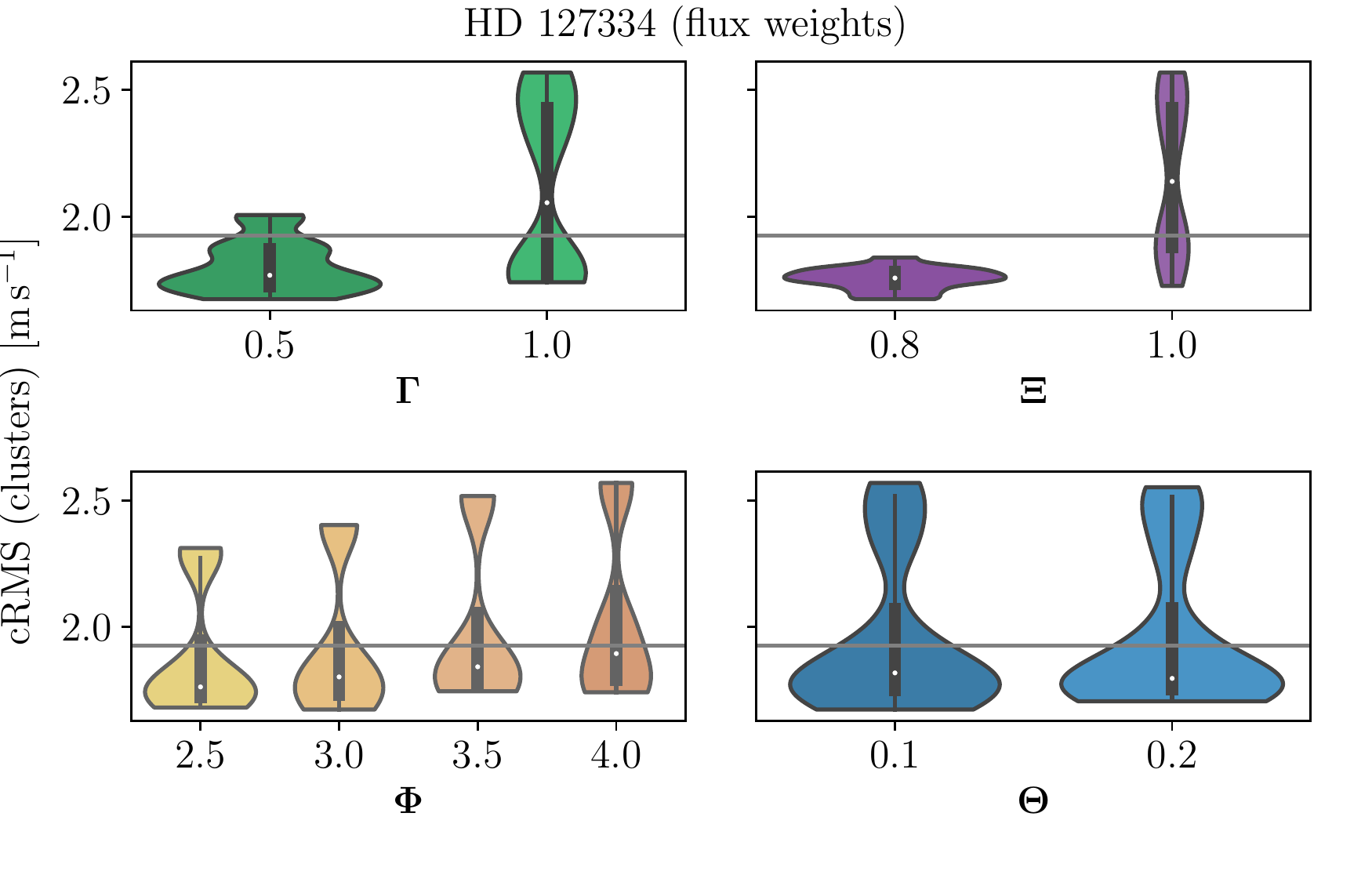}
	\includegraphics[width=\columnwidth]{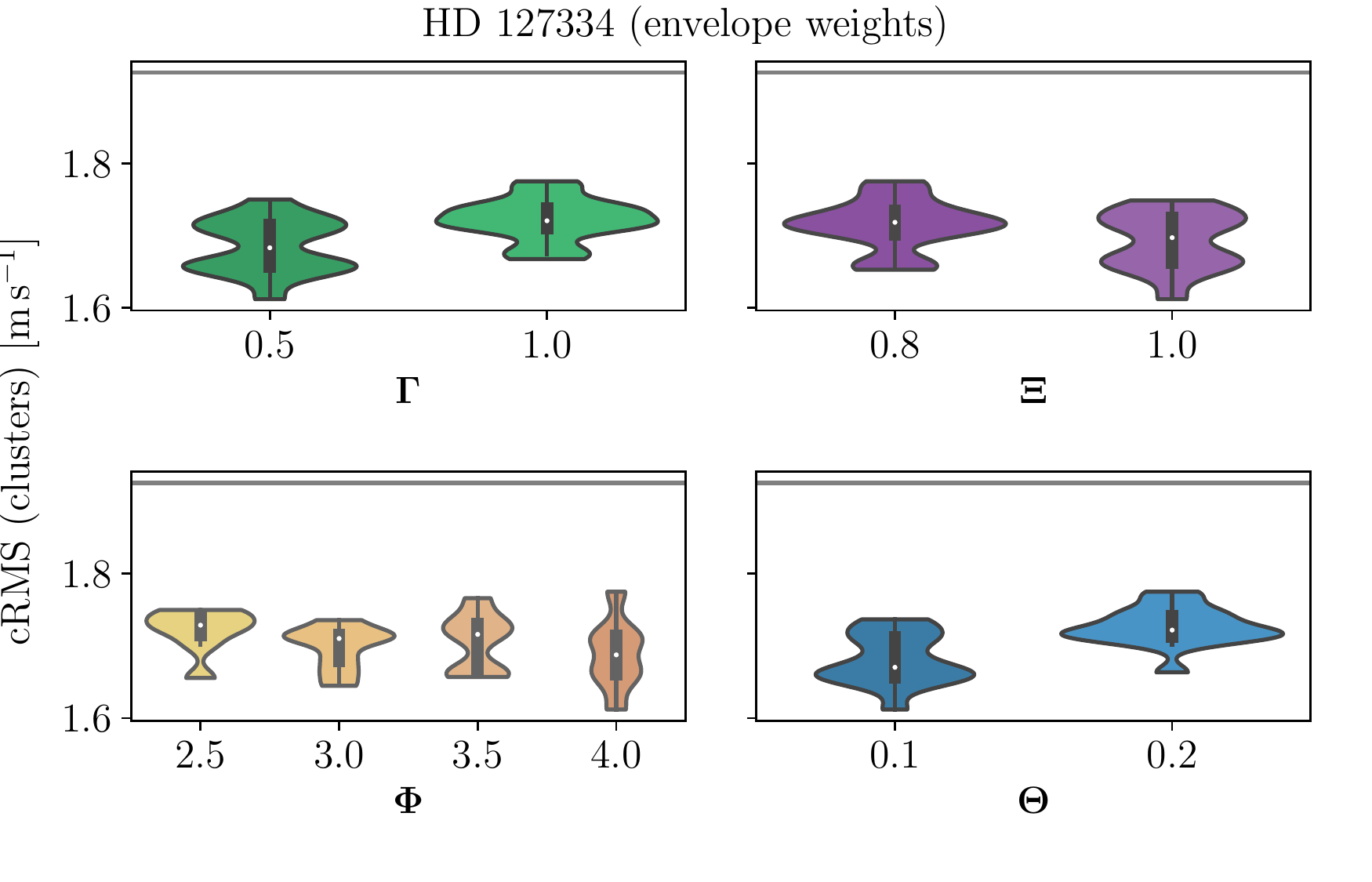}
    \caption{Violin plot showing the cRMS distribution (weighted mean cluster RMS, cRMS) of the 32 LSD RV time series split by the varied parameters (cf. definitions in Section \ref{ss:findparameters}) for HD 127334. The horizontal line shows the equivalent scatter measure for the DRS CCF time series.}
    \label{fig:rms_paramdep_otherstars1}
\end{figure}

\begin{figure}
	\includegraphics[width=\columnwidth]{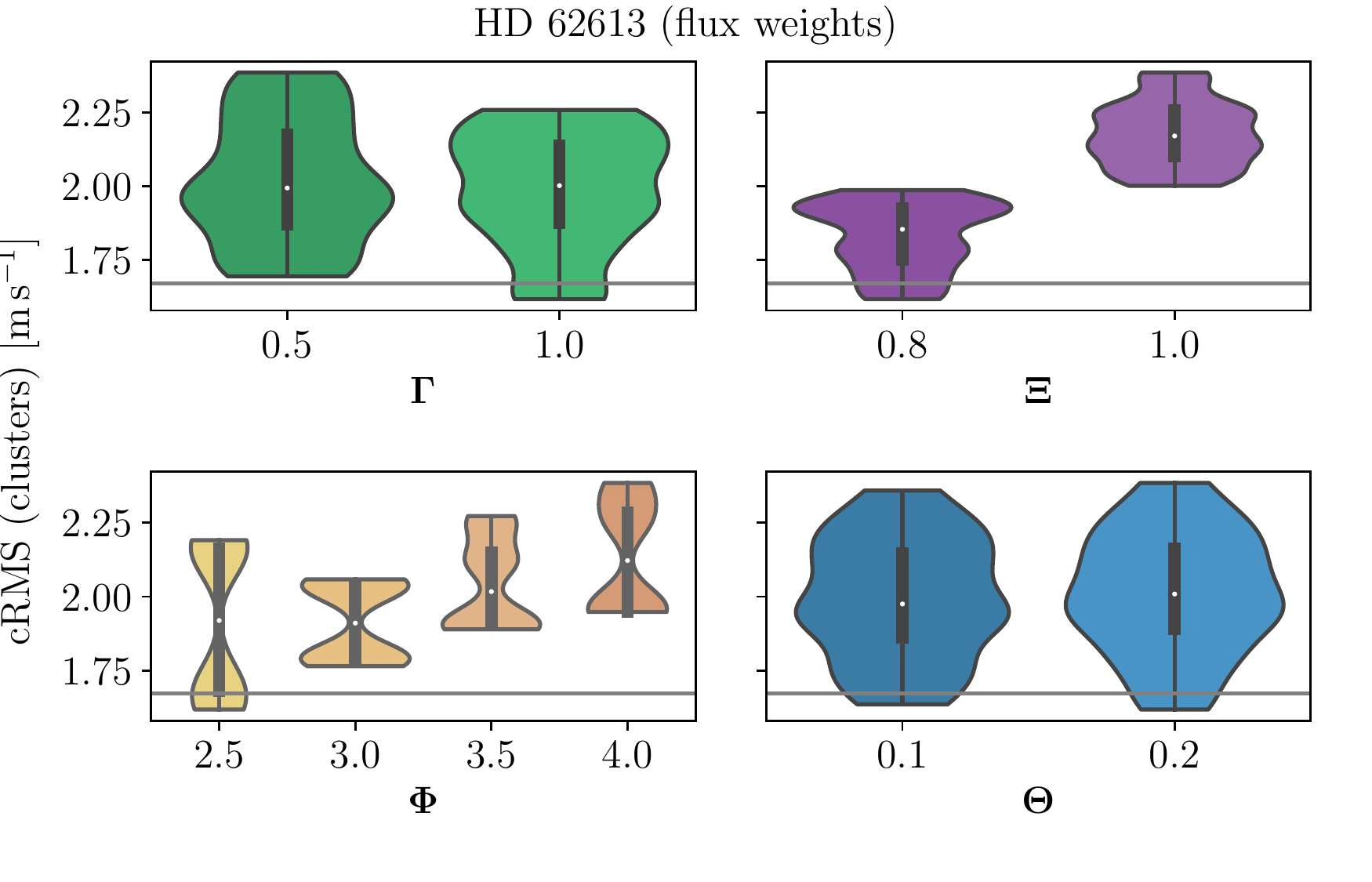}
	\includegraphics[width=\columnwidth]{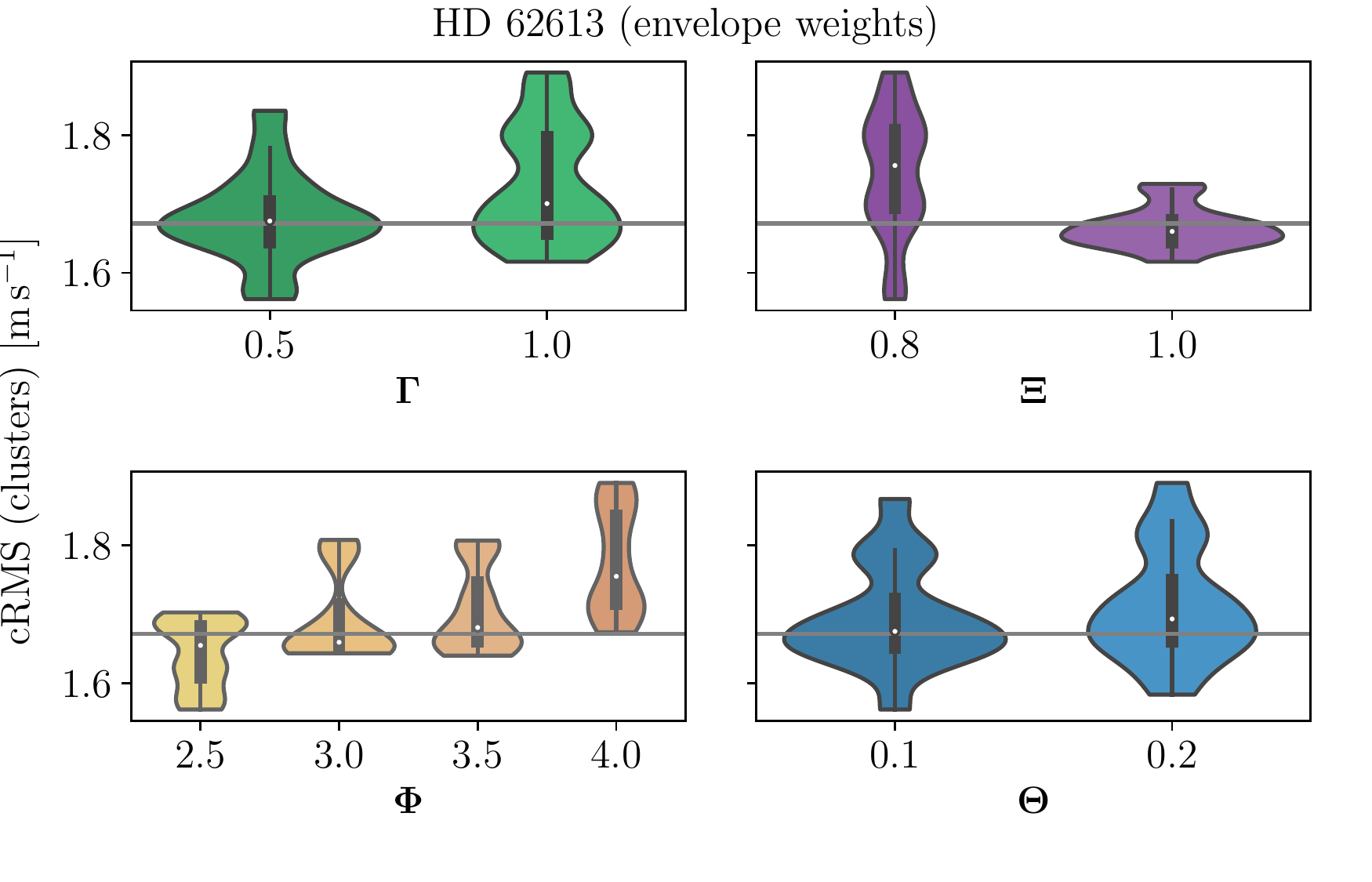}
    \caption{Violin plot showing the cRMS distribution for HD62613.}
    \label{fig:rms_paramdep_otherstars2}
\end{figure}

\bsp	
\label{lastpage}
\end{document}